%% file: main.tex
\documentclass[runningheads, envcountsame, a4paper]{llncs}
\usepackage[pdftex]{graphicx}
\usepackage{epstopdf}

\usepackage{amssymb}
\usepackage{amsmath}
\usepackage{bbm}
\usepackage[caption=false]{subfig}
\usepackage{multicol}
\usepackage{adjustbox}
\usepackage{ascmac}
\usepackage{microtype}
    
\usepackage{times}
\usepackage{xcolor}
\usepackage{soul}
\usepackage[utf8]{inputenc}

\usepackage{bm,amsmath}
\usepackage[ruled]{algorithm2e}
\usepackage{multirow}
\usepackage{color}
\usepackage{amsfonts}
\usepackage{url}

\usepackage{tabularx,tabulary}
\usepackage[misc,geometry]{ifsym}

\begin{document}
\title{Crowdsourcing Evaluation of Saliency-based \\ XAI Methods}
\author{Xiaotian Lu \Letter \inst{1}\orcidID{0000-0002-4118-2301}\and
Arseny Tolmachev\inst{2} \and
Tatsuya Yamamoto\inst{2} \and \\
Koh Takeuchi\inst{1} \and
Seiji Okajima\inst{2} \and
Tomoyoshi Takebayashi\inst{2} \and \\
Koji Maruhashi\inst{2} \and
Hisashi Kashima\inst{1}
}

\authorrunning{Lu et al.}

\institute{Kyoto University, Kyoto, Japan\\
\email{lu@ml.ist.i.kyoto-u.ac.jp, takeuchi@kyoto-u.ac.jp, kashima@i.kyoto-u.ac.jp }\\
\and
Fujitsu Research, Fujitsu Ltd., Kanagawa, Japan\\
\email{$\{$t.arseny,tyamamo,okajima.seiji,\\
takebayashi.tom,maruhashi.koji$\}$@fujitsu.com}}

\maketitle
\sloppy

\input{abstract}
\input{introduction}

\input{related}

\input{proposed}
\input{result}

\input{conclusion}

%\clearpage
\bibliographystyle{splncs04}
%\bibliography{reference}

\if0
\clearpage
\input{memo}
\fi

\end{document}

%% file: abstract.tex
\begin{abstract}

Understanding the reasons behind the predictions made by deep neural networks is critical for gaining human trust in many important applications, which is reflected in the increasing demand for explainability in AI (XAI) in recent years.
Saliency-based feature attribution methods, which highlight important parts of images that contribute to decisions by classifiers, are often used as XAI methods, especially in the field of computer vision.
In order to compare various saliency-based XAI methods quantitatively, several approaches for automated evaluation schemes have been proposed; however, there is no guarantee that such automated evaluation metrics correctly evaluate explainability, and a high rating by an automated evaluation scheme does not necessarily mean a high explainability for humans.
In this study, instead of the automated evaluation, we propose a new human-based evaluation scheme using crowdsourcing to evaluate XAI methods. 
Our method is inspired by a human computation game, "Peek-a-boom", and can efficiently compare different XAI methods by exploiting the power of crowds. 
We evaluate the saliency maps of various XAI methods on two datasets with automated and crowd-based evaluation schemes. 
Our experiments show that the result of our crowd-based evaluation scheme is different from those of automated evaluation schemes. 
In addition, we regard the crowd-based evaluation results as ground truths and provide a quantitative performance measure to compare different automated evaluation schemes.
We also discuss the impact of crowd workers on the results and show that the varying ability of crowd workers does not significantly impact the results.

\keywords{Explainable AI \and Interpretability \and Evaluation \and Crowdsourcing}
\end{abstract}

%% file: introduction.tex
\section{Introduction}

Recent significant advances in AI technologies have introduced innovations in various fields. 
In particular, deep neural networks (DNNs) exhibit remarkable performance in a wide range of real-world applications, such as natural language processing~\cite{collobert2011natural,socher2013recursive}, image classification~\cite{krizhevsky2012imagenet,ciresan2012deep,szegedy2015going,ciregan2012multi}, and human action recognition~\cite{ji20123d,le2011learning}.
DNNs can extract intricate underlying patterns from large and high-dimensional datasets and have reduced the demand for feature engineering.
However, the internal mechanism of DNNs is a black box, i.e., it is difficult to understand the relationships between their inputs and outputs.
In low-risk environments, errors made by DNNs do not have severe impacts; for example, in movie recommendation systems, the impact of making a recommendation error is relatively low.
However, in other fields such as healthcare, a single misdiagnosis can be fatal; therefore, it is essential to explain the predictions.
In regulated industries such as the judicial system and financial markets, a mandate for explanations in addition to model predictions is emerging in legal norms. 
However, most current DNN models are opaque and provide no information about their decision-making process, which has been a significant obstacle to the implementation of AI in essential applications.
Understanding the reasons behind their predictions is critical for gaining human trust in many important applications~\cite{ribeiro2016should}, which is reflected in the increasing demand for explainability in AI (XAI) in recent years.

\begin{figure}[!tb]
    \centering
    \subfloat[Original image]{
        \includegraphics[width=1.38in]{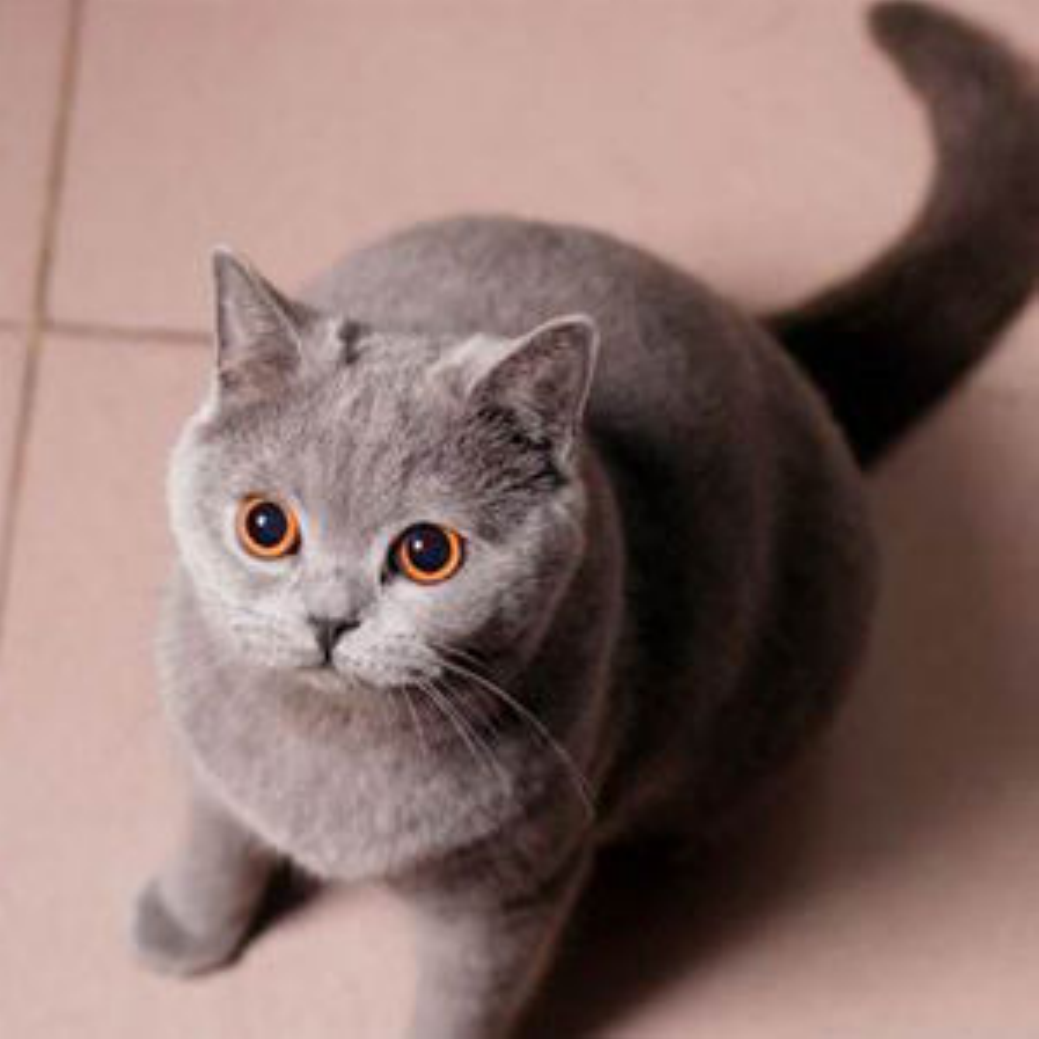}%
            \label{fig:catOrigin}}
    \subfloat[Vanilla Gradients]{
            \includegraphics[width=1.39in]{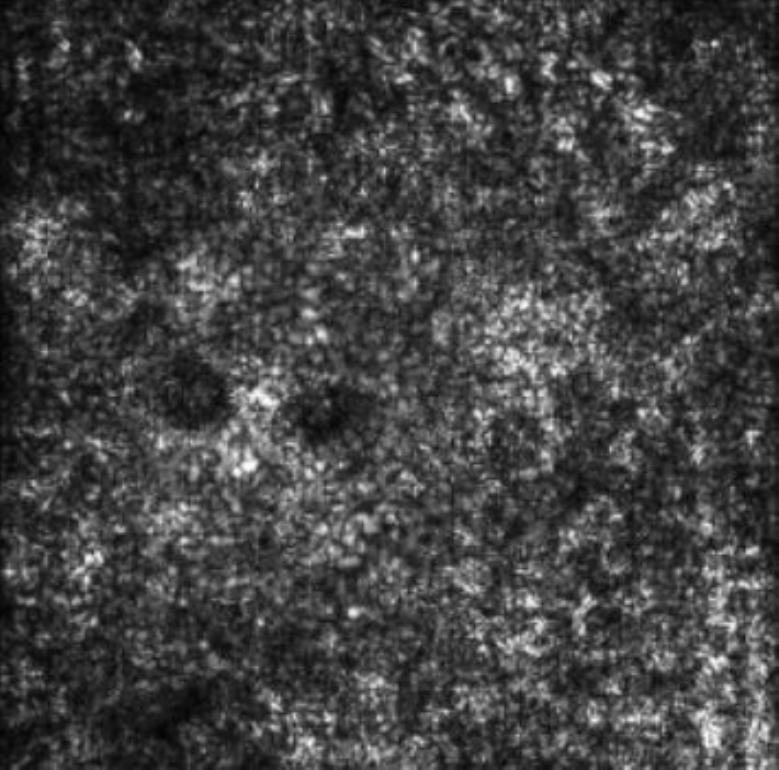}%
            \label{fig:catVanilla}}	
     \subfloat[SmoothGrad]{
            \includegraphics[width=1.38in]{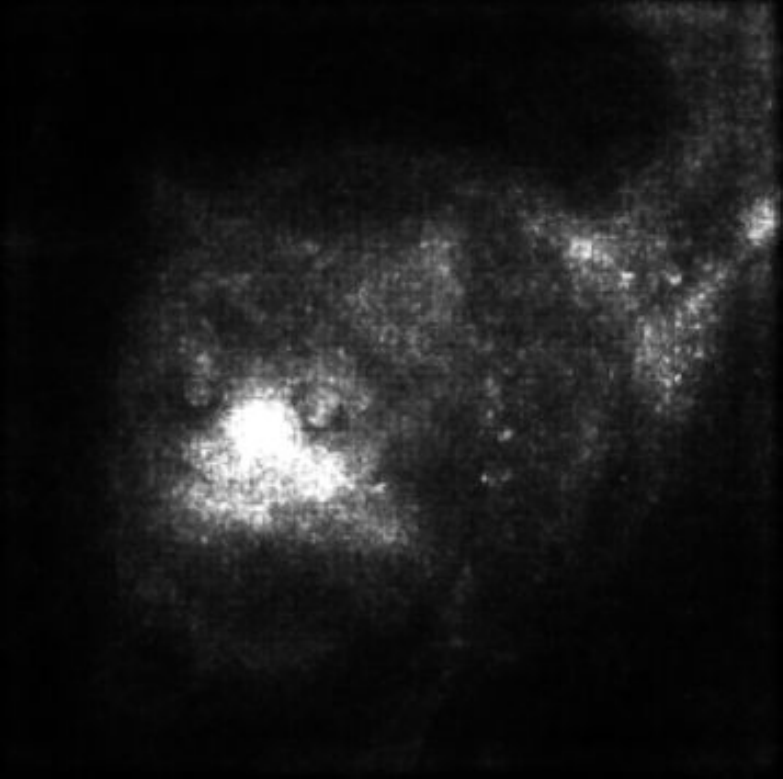}%
            \label{fig:catSmoothgrad}}	\\
    \subfloat[Guided-Backpropagation]{
            \includegraphics[width=1.4in]{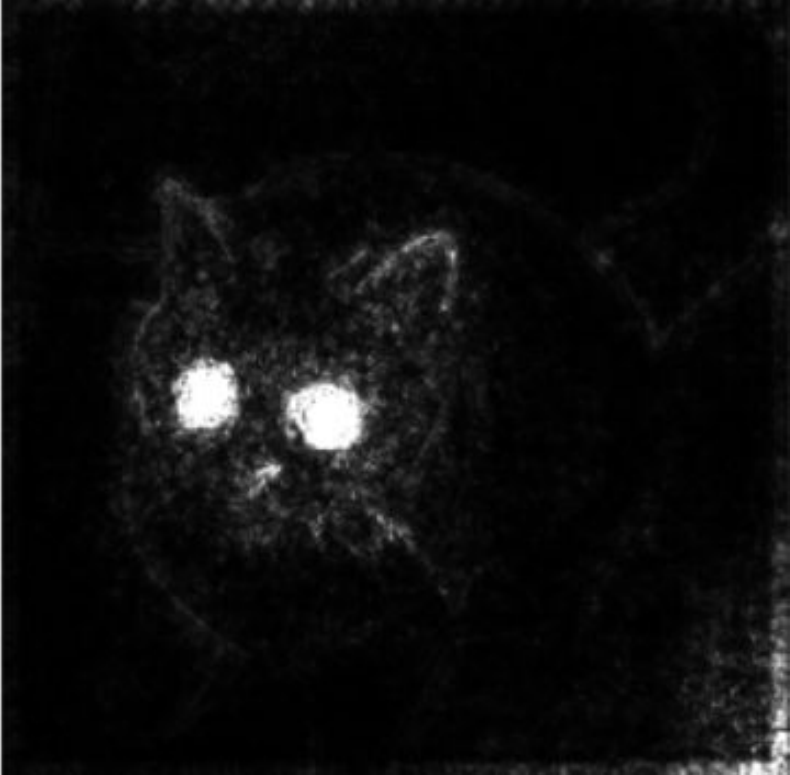}%
            \label{fig:catGuided}}	
    \subfloat[Grad-CAM]{
            \includegraphics[width=1.38in]{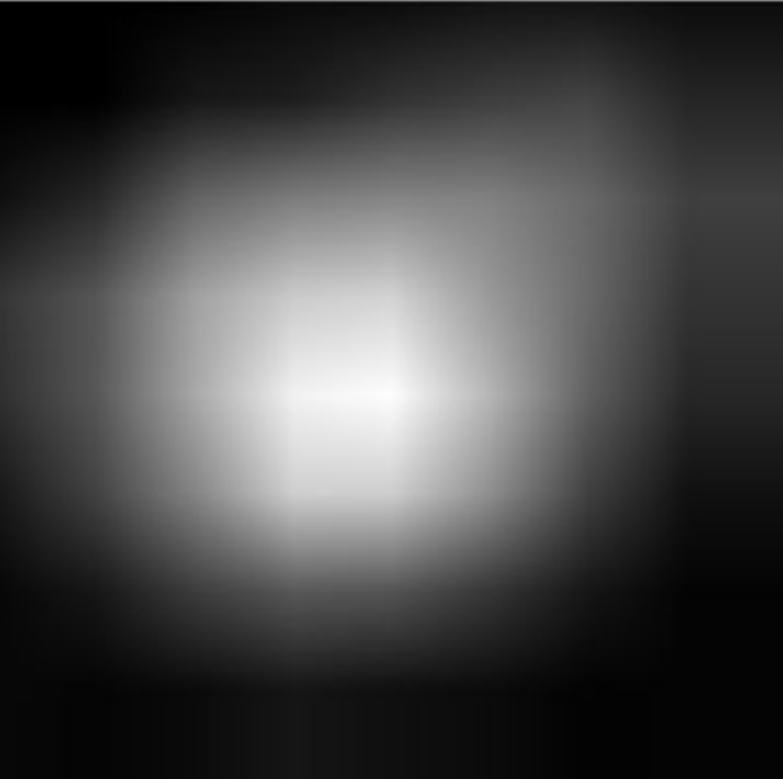}%
            \label{fig:catGradCam1}}	
    \caption{
      Different saliency maps produced by different XAI methods. The bright areas indicate important areas.
    }
    \label{fig:catdiffmap}
\end{figure}

To satisfy the requirements of XAI, various explanation and interpretation methods have emerged, especially for black-box predictions made by already-trained neural networks.
One of the major approaches to this problem involves the estimation of the influence of a subset of input features on the predictions of a model. 
By understanding the important features, the model can be improved, model predictions can be trusted, and undesirable behaviors can be isolated~\cite{hooker2019benchmark}.
For example, in image classification tasks, the generation of saliency maps, which assign an importance measure for each part (or pixel) of an input image, is a major research direction; the representative methods include
Vanilla Gradients~\cite{baehrens2009explain,erhan2009visualizing,simonyan2013deep}, 
SmoothGrad~\cite{smilkov2017smoothgrad}, 
Guided-Backpropagation~\cite{springenberg2014striving}, and
Grad-CAM~\cite{selvaraju2017grad,zhou2016learning}  (Figure~\ref{fig:catdiffmap}).

\if0
Saliency maps help us determine the areas focused on by DNNs, which can make us better understand the decisions of DNNs. 
For example, in Figure~\ref{fig:saliencymap}, we apply a classifier on an airplane image. 
However, the saliency map shows that the blue sky is the area of focus for this classifier. 
Here, this classifier was unaware of airplanes even if it answered correctly.
When we apply the classifier to a pure blue image, the pure blue image will still be considered an airplane.

\begin{figure}[tb]
    \centering
    \subfloat[Original image]{
            \includegraphics[width=1.5in]{figure/airplane.pdf}%
            \label{fig:catOrigin}}
   	\subfloat[Grad-CAM mask]{
            \includegraphics[width=1.5in]{figure/airplaneCAM.pdf}%
            \label{fig:catGradCam}}	\\
    \caption{
     Grad-CAM saliency map shows that the blue sky is the focused area. This is a reasonable strategy for machines to make decisions based on statistical information, but it is probably different from how humans recognize objects. 
    }
    \label{fig:saliencymap}
\end{figure}
\fi

While various XAI methods have been proposed, their evaluation strategy has not been established well, and there is an urgent demand for quantitative measures to answer the question ``Given several XAI methods of a black-box prediction model, which one yields the best interpretations?"
Several automated evaluation schemes have been proposed~\cite{hooker2019benchmark,samek2016evaluating,nguyen2020model}; they usually delete or replace pixels that are said important by an XAI method, and check the deterioration in the prediction performance.
However, as pointed out by a recent research~\cite{narayanan2018humans}, high interpretability for machines does not imply the same for humans. 
A machine may recognize an object based on its relation to the background rather than the object itself. 
For example, the background of an image of an airplane is often the sky. 
This is a reasonable strategy for machines to make decisions based on statistical information, but it is probably different from how humans recognize objects. 
After all, interpretability for humans can ultimately only be evaluated by humans.

\begin{figure}[!tb]
\captionsetup[subfigure]{labelformat=empty}
    \centering
         \subfloat[Exposure rate of 5\%]{
            \includegraphics[width=1.6in]{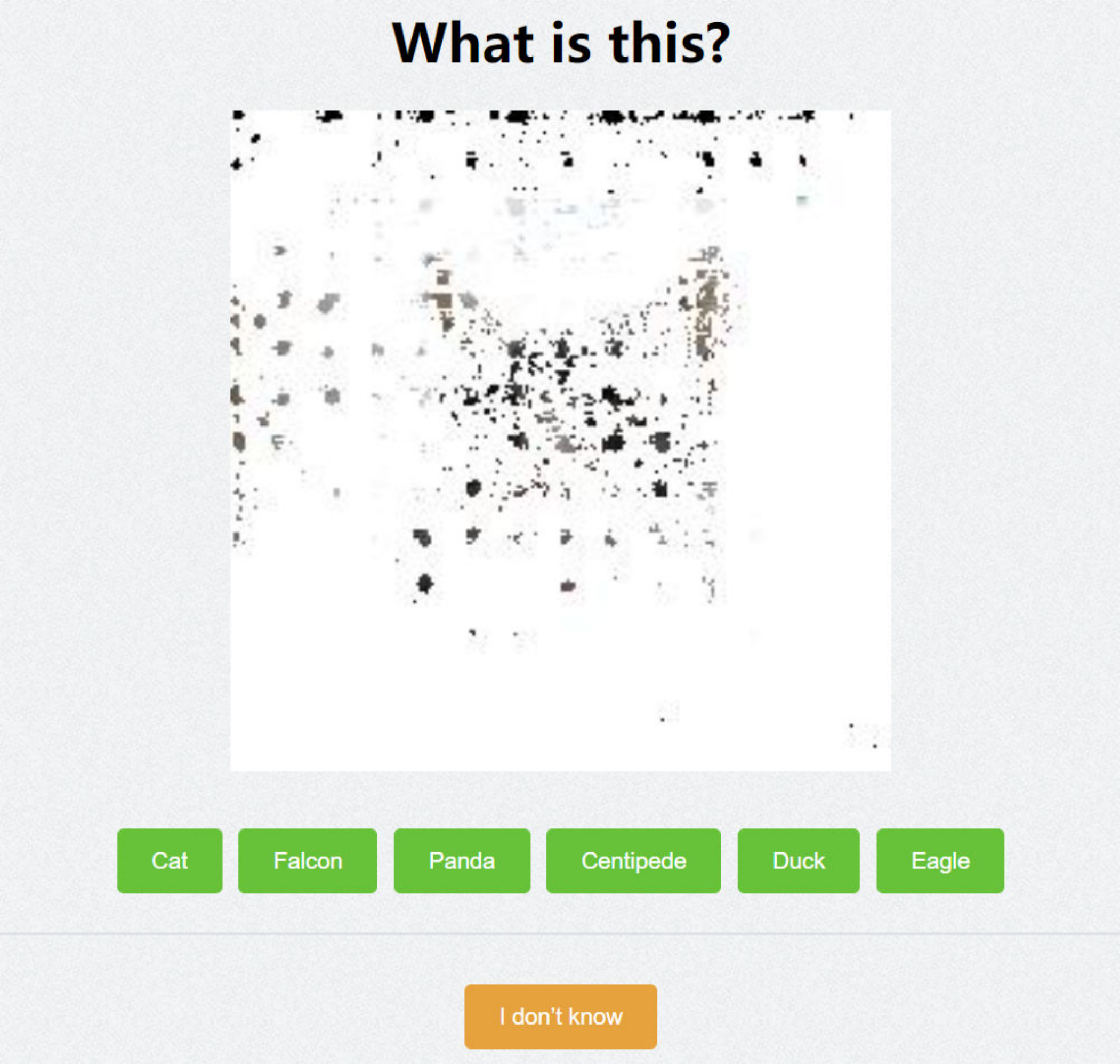}
            }
            \hspace{10mm}
         \subfloat[Exposure rate of 10\%]{
            \includegraphics[width=1.6in]{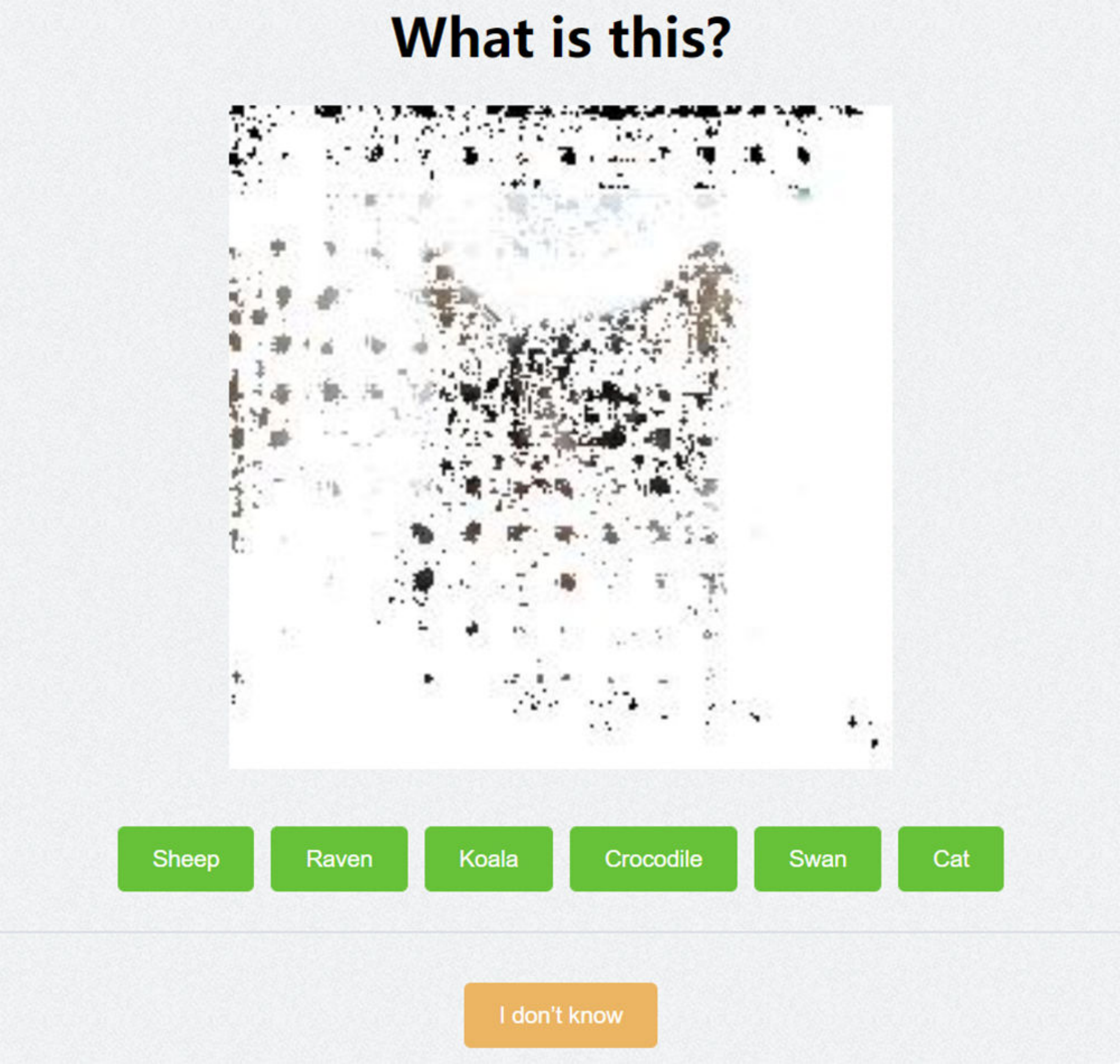}
            } \\
         \subfloat[Exposure rate of 15\%]{
            \includegraphics[width=1.6in]{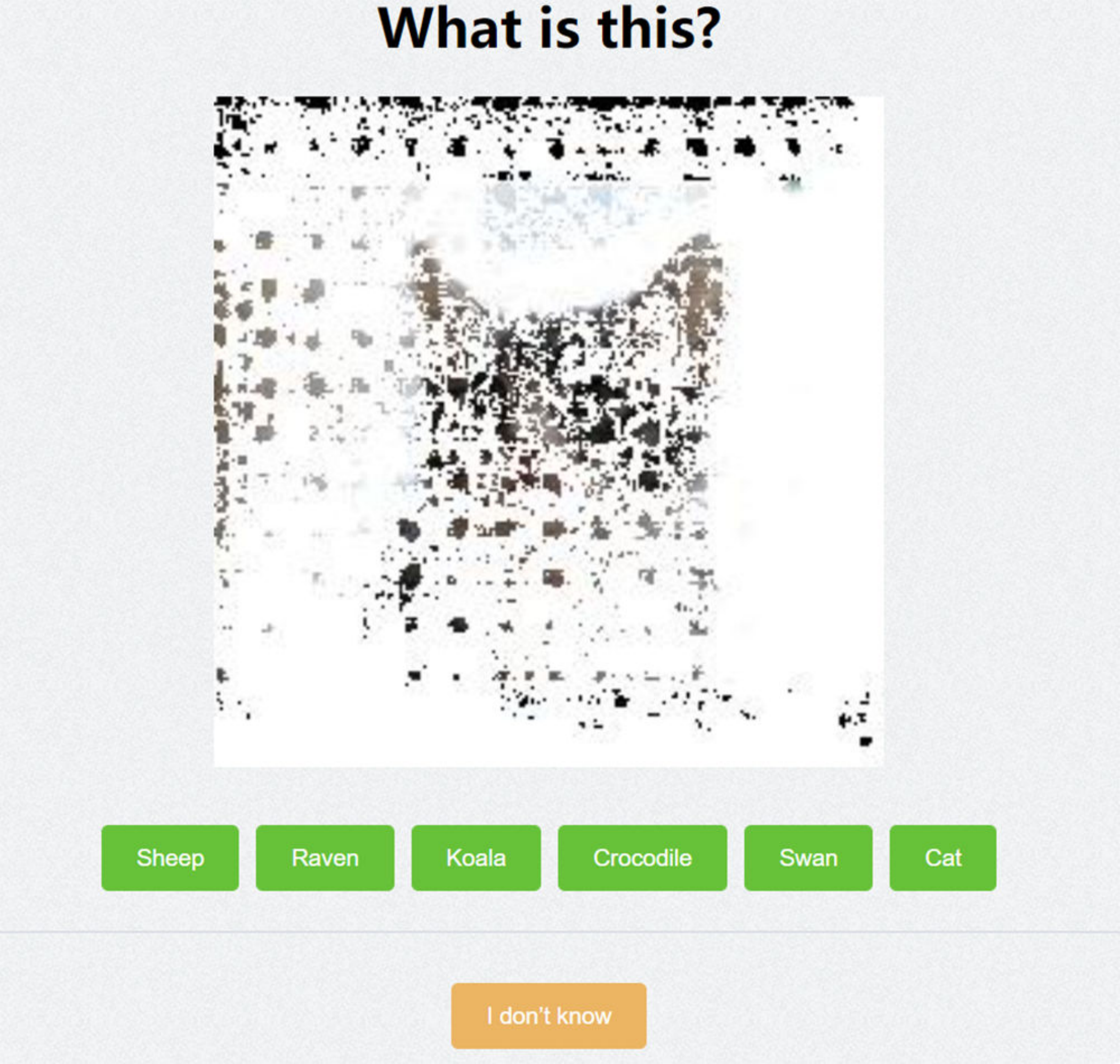}
            }
            \hspace{10mm}
         \subfloat[Exposure rate of 30\%]{
            \includegraphics[width=1.6in]{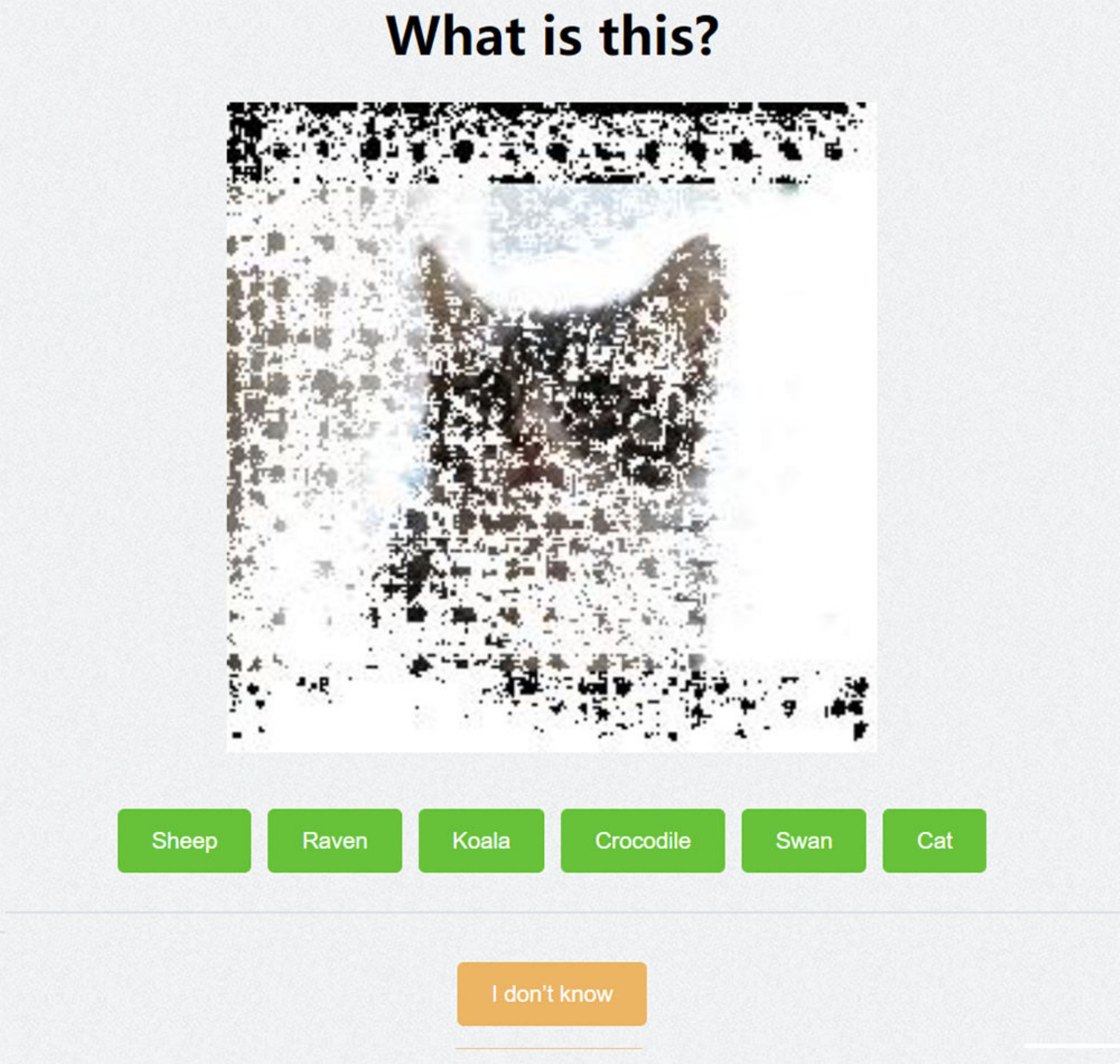}
            }
    \caption{
    Example of how the proposed evaluation interface gradually reveals an image to crowd workers at different exposure rates (5\%. 10\%, 15\%, and 30\%).
    }
    \label{fig:gradshow}
\end{figure}

In this study, we perform human-based evaluation of XAI methods by using crowdsourcing.
Our XAI evaluation approach is based on a human computation game.
In human computation~\cite{law2011human}, some approaches embed human intelligence tasks into games, usually referred to as game with a purpose (GWAP).
Peek-a-boom is a type of GWAP with two players named Peek and Boom; Boom reveals a part of a given image and Peek guesses the image object from the revealed part.
We use a Peek-a-boom-based XAI evaluation in which a human plays the Peek, and an XAI method plays the Boom instead of another human.

We implement a crowd-based evaluation interface (Figure~\ref{fig:gradshow})\footnote[1]{Our crowdsourcing evaluation interface can be tried at \url{https://17bit.github.io/crowddemo/index.html} .}, recruit crowd workers for executing evaluation tasks of four popular XAI methods on two real datasets, and compare the results with those by four automated evaluation schemes.

The results show that the proposed scheme gives  different evaluations from the automated evaluation schemes.
Subsequently, We consider the proposed crowd-based scheme as ground truths and evaluate the automated evaluation schemes in terms of interpretability for humans.
Finally, we analyze the ability of crowd workers and find that, even if their ability may vary considerably, the final results are not significantly affected. 

The contributions of this study are summarized as follows:\\
\noindent
1) We propose a new crowd-based evaluation scheme for XAI methods. \\
2) We experimentally investigate the difference between automated and crowd-based evaluation schemes.\\
3) We provide a performance measure for automated evaluation schemes based on their similarity to the proposed crowd-based evaluation scheme. \\
4) We examine the impact of the number and ability of crowd workers on the results.

%% file: related.tex
\section{Related Work}
We briefly review the XAI methods used in this study, automated evaluation schemes that can automatically evaluate the XAI methods, and existing crowd-based evaluation schemes.

\subsection{XAI methods}
Most of the existing XAI methods attribute the output of a pre-trained neural network to a part of its input.
For a multi-class classification problem with $C$ classes, let $f: \mathbb{R}^D\to \mathbb{R}^C $ be a pre-trained neural network which takes an input feature vector $x\in \mathbb{R}^D$ and output a vector representing the degree of classification into each class.
A typical XAI method provides a \textit{saliency map} 
$s: \mathbb{R}^D\to \mathbb{R}^D$ that maps an input feature vector to a vector whose $d$-th element indicates the importance of the $d$-th feature of input $x$~\cite{adebayo2018sanity}.

{\em Vanilla gradient}~\cite{baehrens2009explain,erhan2009visualizing,simonyan2013deep}
is the most basic method to create a saliency map as 
$s^{\text{Vanilla}} (x) = \frac{\partial f}{\partial x}$, which quantifies the influence of a small change in each input dimension on the output of the network.
One of the drawbacks of this method is that the results are sensitive to the noise in the original image (Figure \ref{fig:catVanilla}).

{\em SmoothGrad}~\cite{smilkov2017smoothgrad}
is an improved version of the vanilla gradient which estimates the gradient more robustly.
It takes the average of the vanilla gradient over perturbed inputs, $ s^\text{Smooth}(x)=\frac{1}{N} \sum_{i=1}^{N} s^\text{Vanilla}(x +\Delta_i)$,
where the perturbation of the input $\Delta_i$ is sampled by a Gaussian distribution,
which results in clearer saliency maps than the vanilla gradient (Figure \ref{fig:catSmoothgrad}).

{\em Guided-Backpropagation}~\cite{springenberg2014striving} gives the contribution of an input dimension of a neuron to the output by distributing the output back to the input.
For better interpretability, it back-propagates the output of ``active" ReLU units,
which highlights important edges in images (Figure \ref{fig:catGuided}).

{\em Grad-CAM}~\cite{selvaraju2017grad} focuses on the last convolution layer of a CNN (Convolutional Neural Network) and visualizes the globally-average-pooled gradients
In contrast with the previously mentioned XAI methods, only Grad-CAM relies on both gradients and feature maps of the convolutional layer.
The results focus more on important ``areas" rather than edges as shown in Figure \ref{fig:catGradCam1}.
A recent research shows that these saliency maps based on only the gradients will not change greatly even if the parameters of DNN model are randomized~\cite{adebayo2018sanity}. 
However, saliency map of GradCAM is different from saliency maps based on gradients. Our experiments also show that GradCAM performs the best in our crowd-based evaluation scheme.

\subsection{Automated evaluation schemes for XAI methods}
\label{subsec:xai_methods}

Most automated evaluation schemes work by modifying (either train or test) data, and compare the differences in the prediction performance of models.
Hooker et al.~\cite{hooker2019benchmark} proposed several automated evaluation schemes for XAI methods.
In their studies, they argued that by removing data features from the training set, better evaluation robustness can be archived in comparison to schemes that modify the test set.

In the Remove and Retrain (ROAR) scheme, the top-ranked pixels given by XAI methods are removed from the images in the training dataset, a new model is trained on the modified training set, and the resulting model is evaluated on the non-modified test dataset. The ROAR scheme was mainly compared to the Keep and Retrain (KAR) scheme, in which the bottom-ranked pixels were removed from the training set. 
It should be noted that Hooker et al. did not perform comparisons to crowd-based evaluation schemes.

In our experiments (Section 4), we also test two schemes that change the test set while leaving the training set unchanged, namely, Remove and Evaluate (ROAE) and Keep and Evaluate (KAE)~\cite{samek2016evaluating,nguyen2020model}.
Table~\ref{tab:xai_eval_methods} summarizes the comparison between the automated evaluation schemes.

\begin{table}[!tb]
    \centering
        \caption{Dataset split modification of automated evaluation schemes. ``top" indicates that the top-ranked pixels are modified first, ``bottom" implies that the bottom-ranked pixels are removed first, and `-' means that the dataset split is used without modification. 
        }
    \begin{tabular}{c|cc}
         ~~XAI method~~ & ~~~~test set~~~~ & ~~~~training set~~~~ \\ % What are "train" and "test"?
         \hline
         ROAR~\cite{hooker2019benchmark} & - & top \\
         KAR~\cite{hooker2019benchmark} & - & bottom \\
         ROAE~\cite{samek2016evaluating,nguyen2020model} & top & - \\
         KAE~\cite{samek2016evaluating,nguyen2020model} & bottom & - \\
    \end{tabular}
    \label{tab:xai_eval_methods}
\end{table}

\subsection{Crowd-based evaluation schemes for XAI methods}
Several crowd-based evaluation schemes have been proposed to measure the ability of XAI methods.
Hutton et al.~\cite{hutton2012crowdsourcing} used crowdsourcing to assess the explanations for supervised text classification. 
Crowd workers were asked to compare human-generated and XAI method-generated explanations and indicate which they preferred and why. 
 Selvaraju et al.~\cite{selvaraju2017grad} and Jeyakumar et al.~\cite{jeyakumar2020can} asked crowd workers to choose better explanations directly. Similarly, Can et al.~\cite{can2018ambiance} asked crowd workers to rate the saliency maps of Grad-CAM on the visual characteristics of venues.

Doshi-Velez and Kim~\cite{doshi2017towards} concluded that there are three different types of crowd-based evaluation schemes:
1) binary forced choices in which humans are presented with pairs of explanations and choose better ones, 
2) forward simulation/prediction in which humans are presented with an explanation and an input, and simulate the output of the model, and 3) counterfactual simulation in which humans are presented with an explanation, an input, and an output, and tell what must be changed to change the prediction to a desired output. 

Our proposed scheme is similar to none of the aforementioned schemes. 
We do not directly show the saliency maps to workers nor force them to make a binary choice; instead, we transform into a simpler task, which makes it easier for workers to make objective choices that are less dependent on subjective judgments of workers.

%% file: proposed.tex
\section{Proposed crowd-based evaluation scheme for XAI methods}
\begin{figure}[!tb]
    \centering
         \includegraphics[width=3.0in]{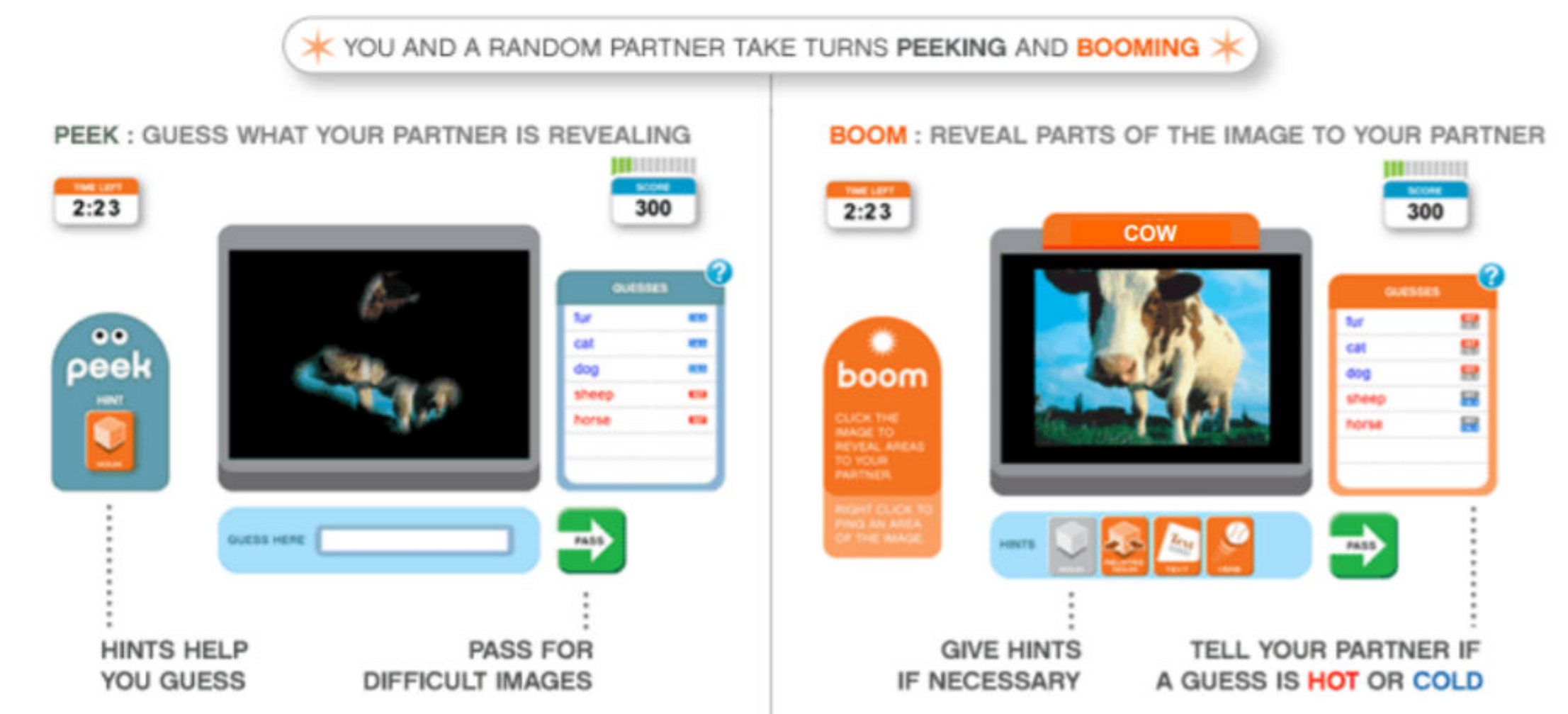}
    \caption{
      Interface of Peek-a-boom human computation game~\cite{von2006peekaboom}. Peek sees the left screen, and Boom sees the right one. Boom determines which parts are exposed to Peek so that Peek correctly guess the image content (that is a cow in this example.)
    }
    \label{fig:Peek-a-boom}
\end{figure}

We propose a crowd-based evaluation scheme for XAI methods based on Peek-a-boom~\cite{von2006peekaboom} which is an online human computation game as shown in Figure~\ref{fig:Peek-a-boom}.
As suggested by the name, this cooperative game has two players, namely, "Peek" and "Boom." 

Peek starts with a blank screen, while Boom starts with an image and a word related to it.
At each round of the game, Boom can specify a small area in the image and reveals the area to Peek, and Peek enters a guess of the word on the basis of the revealed parts.
The both players get more points when Peek correctly answers the word earlier; therefore, Boom has an incentive to reveal only the areas of the image necessary for Peek to guess the correct word.

We use a Peek-a-boom style Web interface, in which a crowd worker plays the Peek, and an XAI method plays the Boom instead of another human.
In the web interface, crowd workers are asked to perform an image classification task, i.e., assigning a label to an image from a set of labels. 
First, we reveal a small percentage of image pixels with an option to reveal more if it is impossible to assign a label with confidence.
We show a correct label, several randomly selected wrong labels, and an ``I don't know" button. 
If the worker cannot provide a confident answer, they can select ``I don't know"; then, more parts of the image will be revealed. 
In the case of an incorrect answer, more parts will be revealed as well. 
Once the worker selects the right answer, or ``I don't know" is selected with a fully-shown image, we give the worker a new image. 
For a given image, XAI methods rank the pixels in descending order of their importance.
For each crowd worker, our crowdsourcing interface starts from an almost blank image (i.e., exposure rate $=0.05$); gradually, the pixels are revealed in order of importance (i.e., increase the exposure rate) (Figure \ref{fig:gradshow}).
At each exposure rate $r \in [0,1]$, the crowd worker is asked to guess the object in the image, typically in terms of multiple-choice questions.
We consider that if an XAI method is "interpretable" enough, the crowd worker can correctly answer the question at a small exposure rate $r$.

\begin{figure}[!tb]
\centering
\renewcommand{\arraystretch}{5}
\centering
\begin{tabular}{cccccccc}
 ~&\multicolumn{7}{c}{\textbf{Exposure Rate}} \vspace{-10mm} \\
 ~ &
 \textbf{5\%} &
 \textbf{10\%} &
 \textbf{15\%} &
 \textbf{20\%} &
 \textbf{30\%} &
 \textbf{50\%} &
 \textbf{75\%} \\

 \rotatebox{90}{\textbf{Grad-CAM}} &
 \includegraphics[width=0.62in]{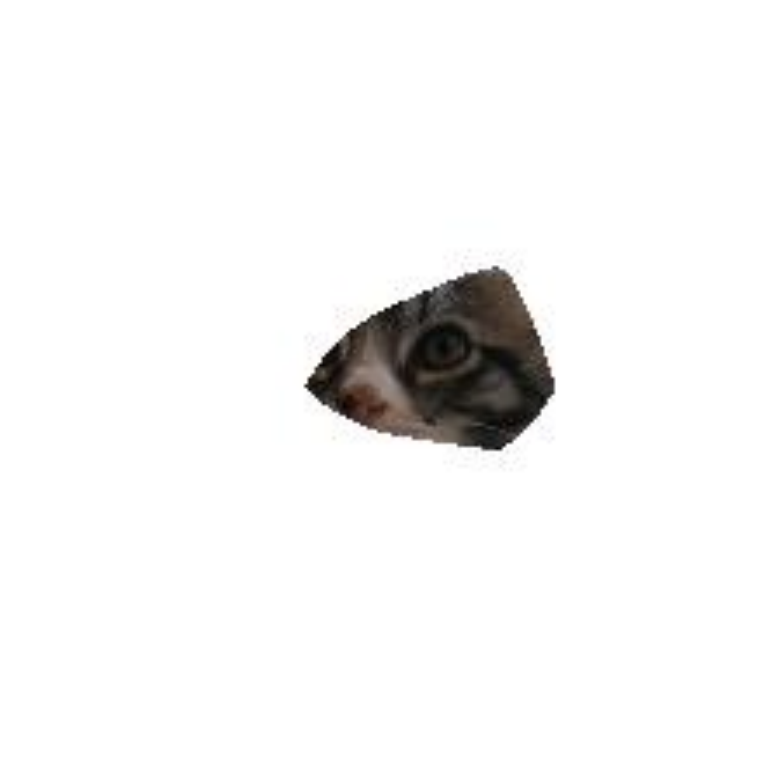} &
 \includegraphics[width=0.62in]{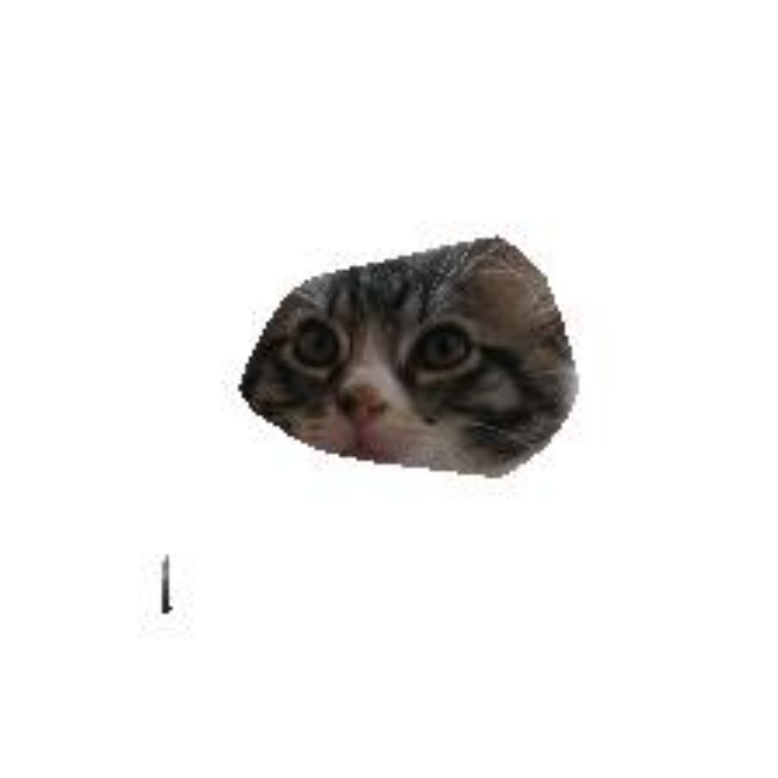} &
 \includegraphics[width=0.62in]{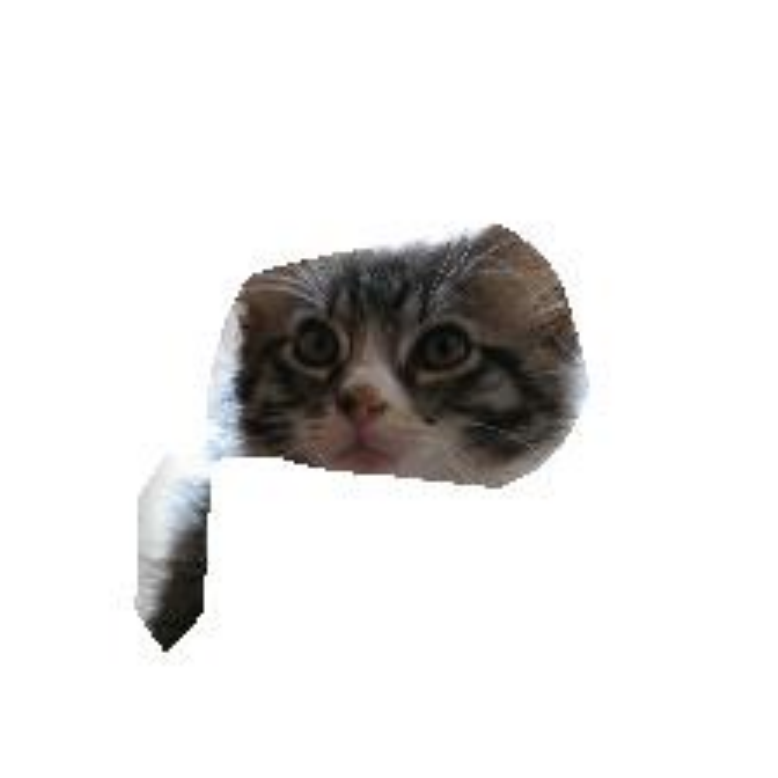} &
 \includegraphics[width=0.62in]{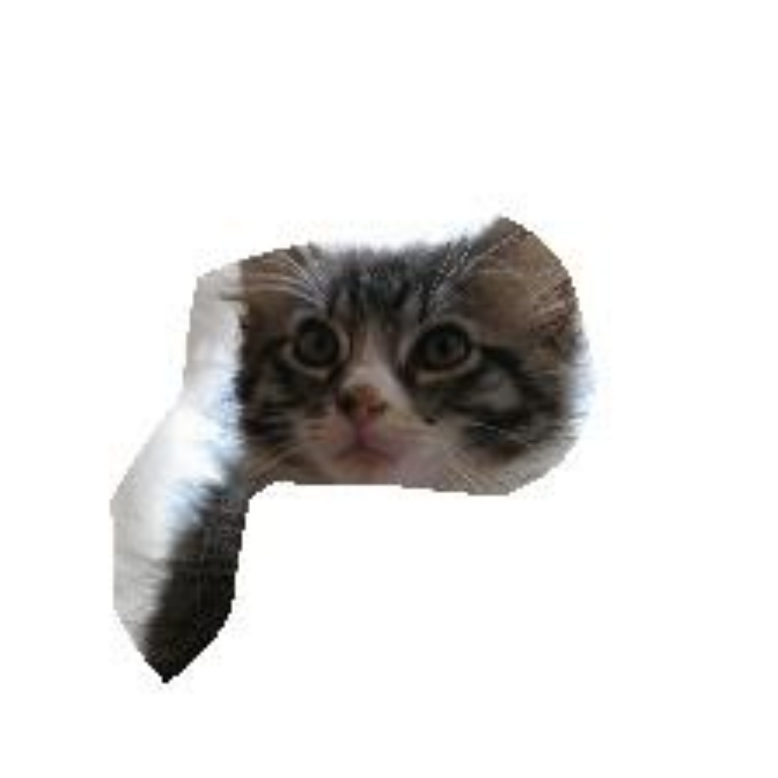} &
 \includegraphics[width=0.62in]{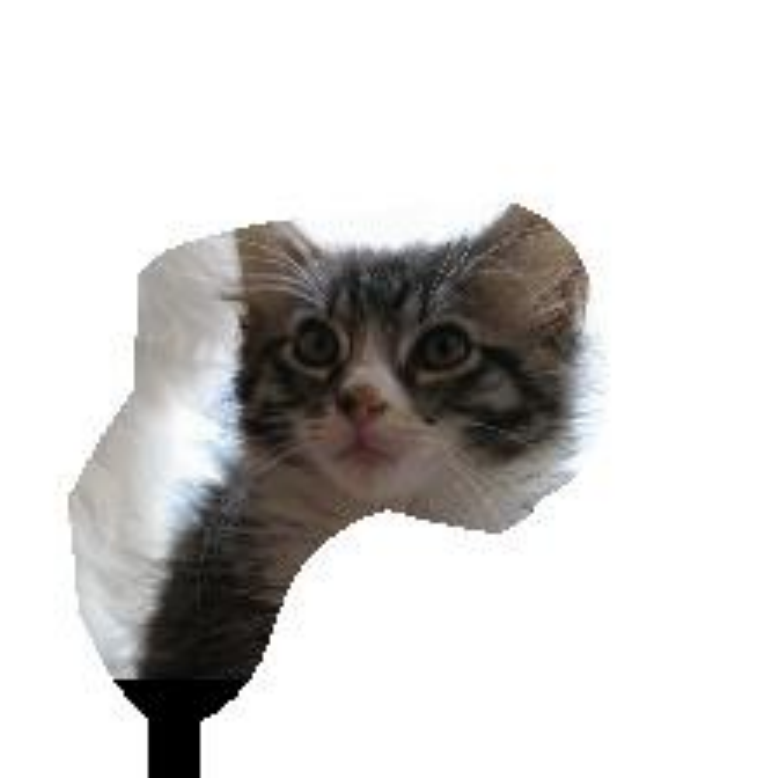} &
 \includegraphics[width=0.62in]{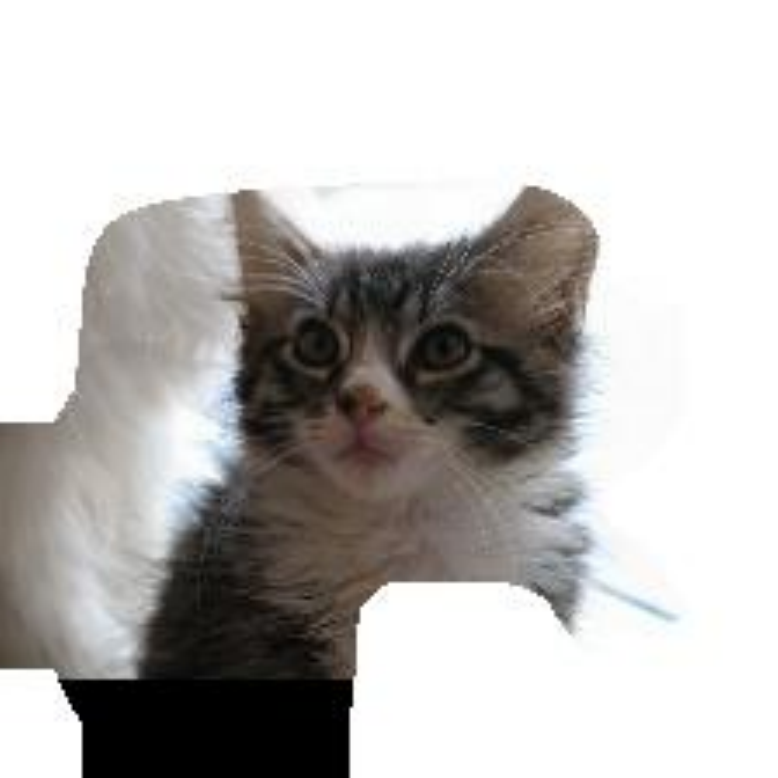} &
 \includegraphics[width=0.62in]{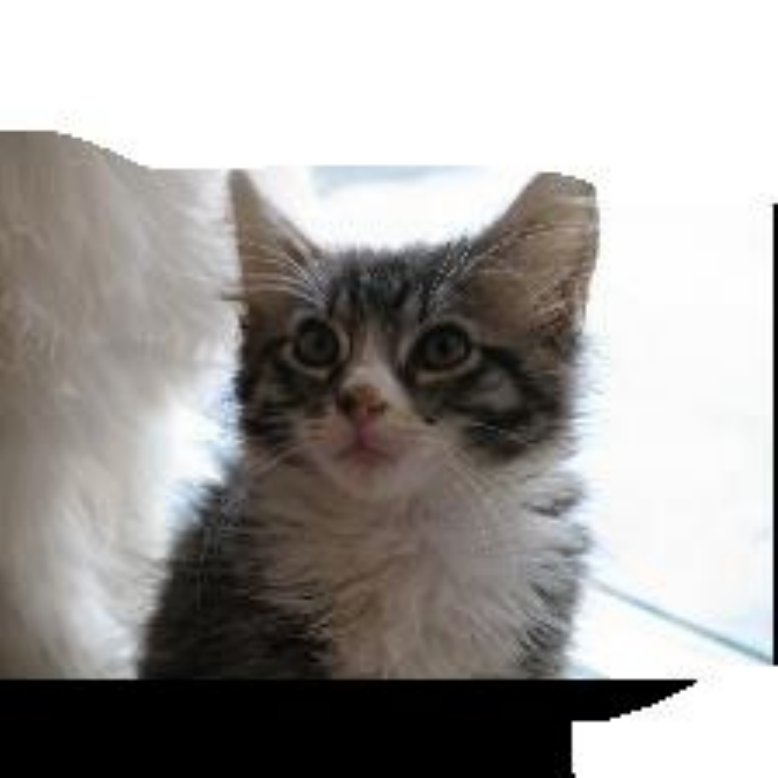}\\

 \rotatebox{90}{~\textbf{Guided-BP}} &
 \includegraphics[width=0.62in]{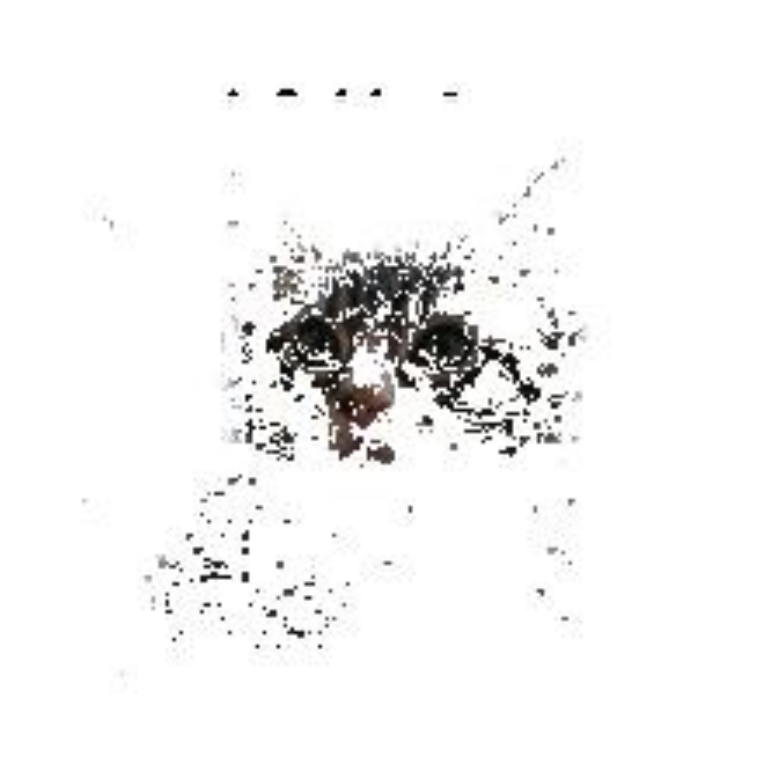} &
 \includegraphics[width=0.62in]{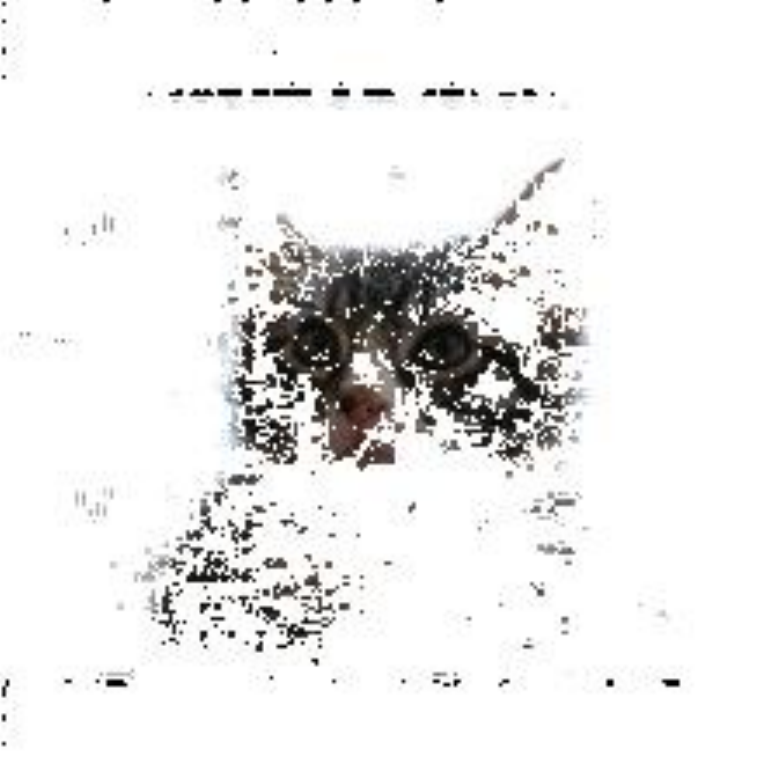} &
 \includegraphics[width=0.62in]{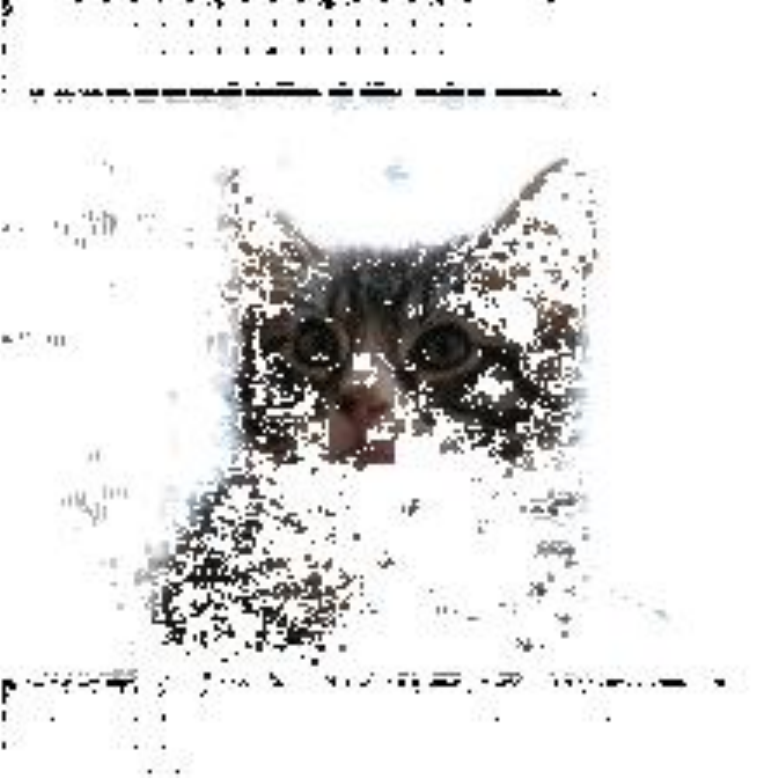} &
 \includegraphics[width=0.62in]{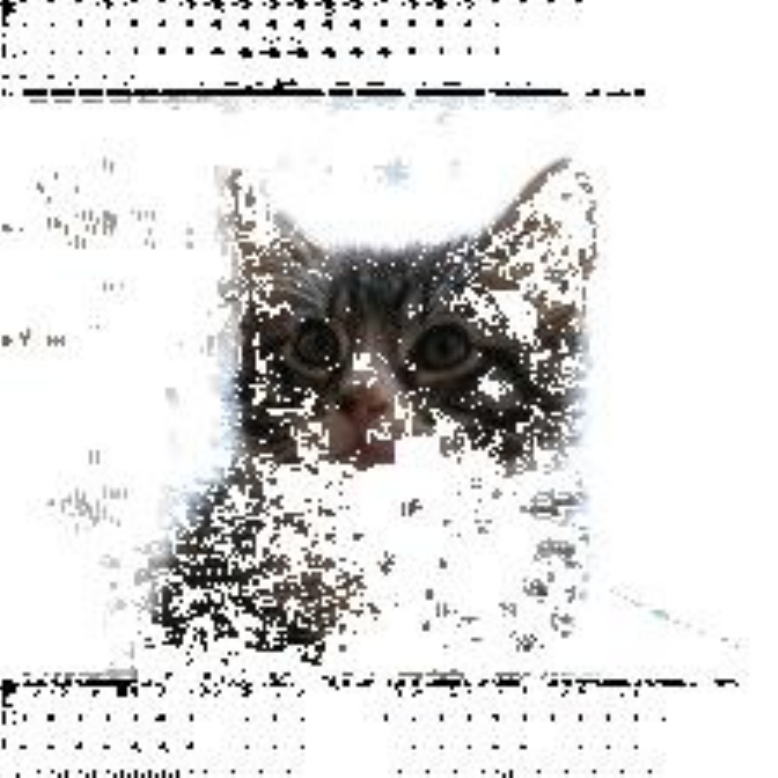} &
 \includegraphics[width=0.62in]{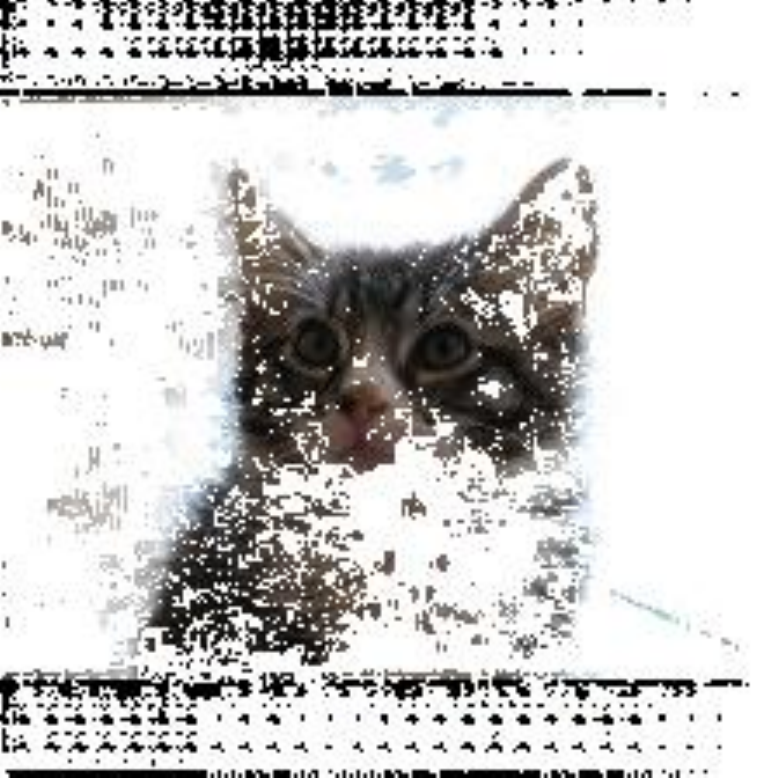} &
 \includegraphics[width=0.62in]{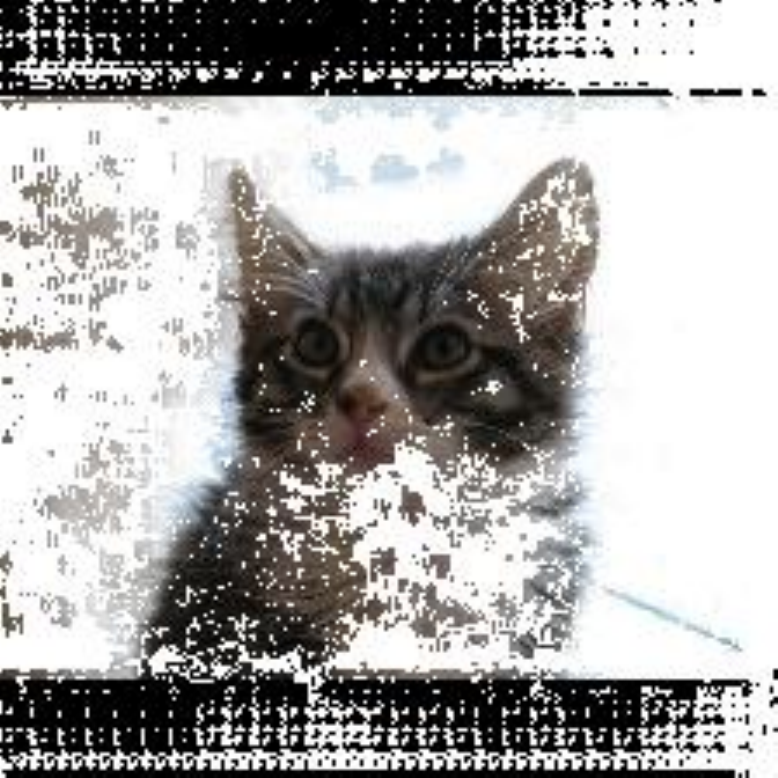} &
 \includegraphics[width=0.62in]{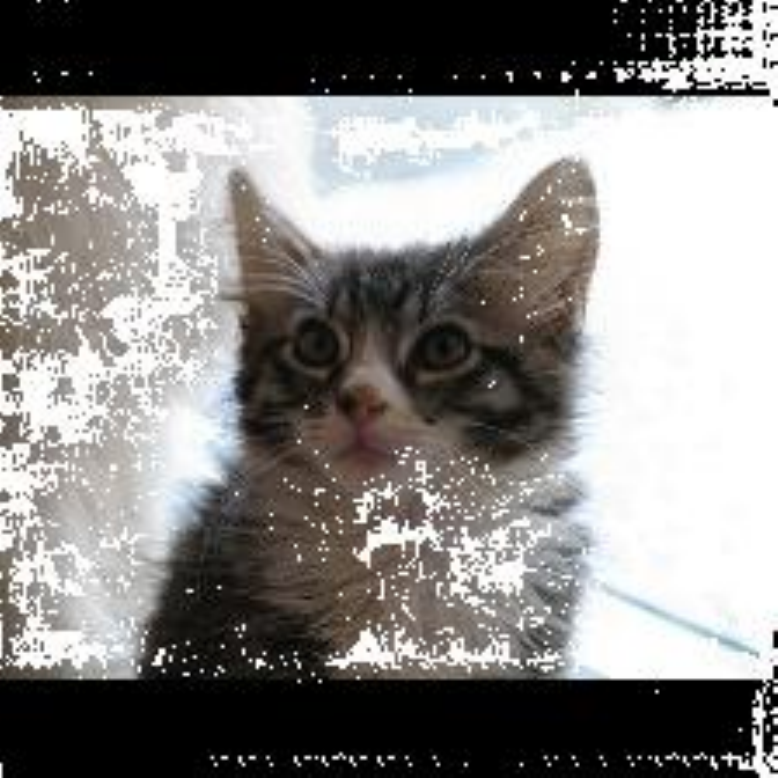}\\
 
  \rotatebox{90}{~\textbf{SmoothGrad}} &
 \includegraphics[width=0.62in]{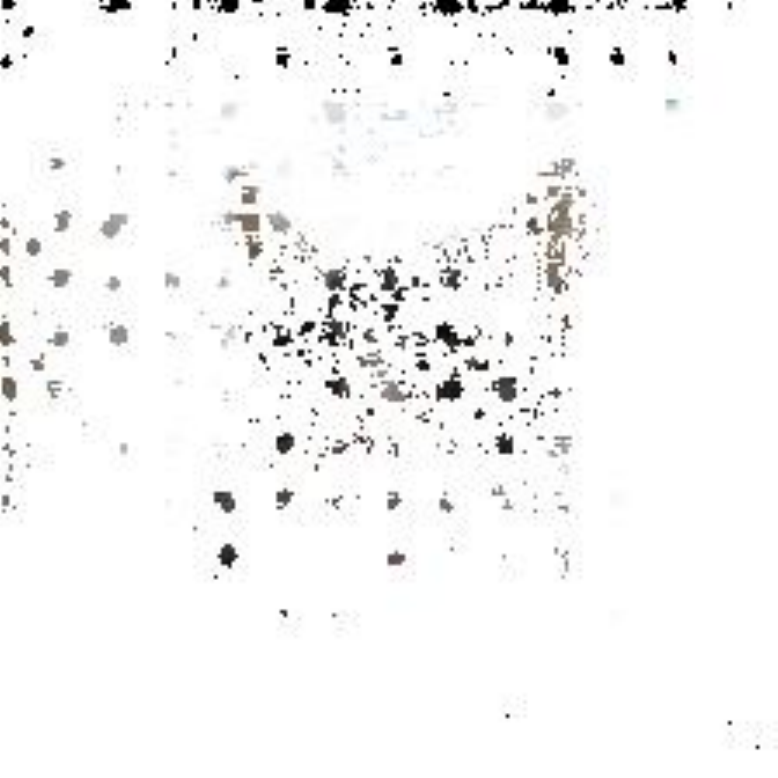} &
 \includegraphics[width=0.62in]{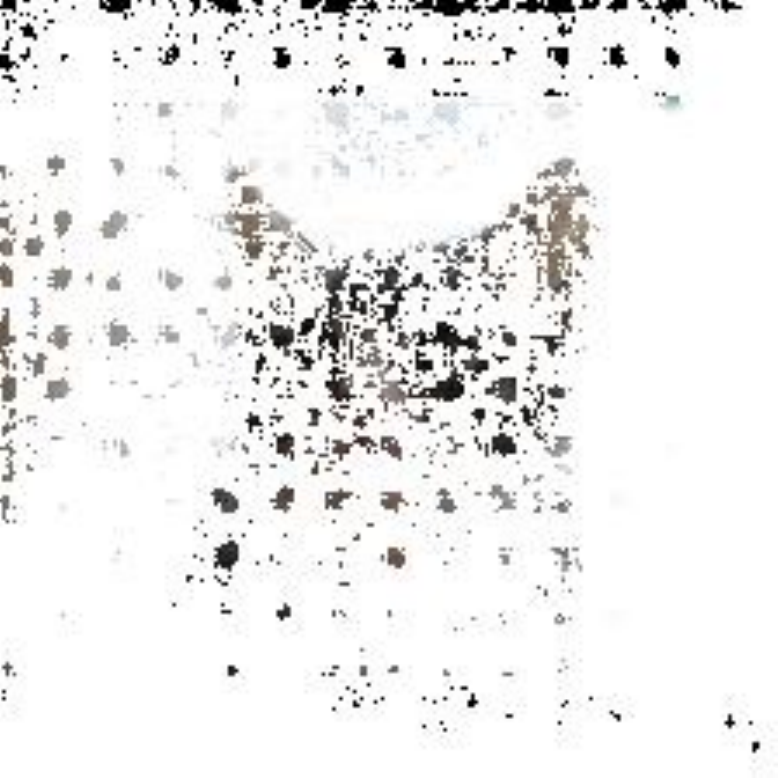} &
 \includegraphics[width=0.62in]{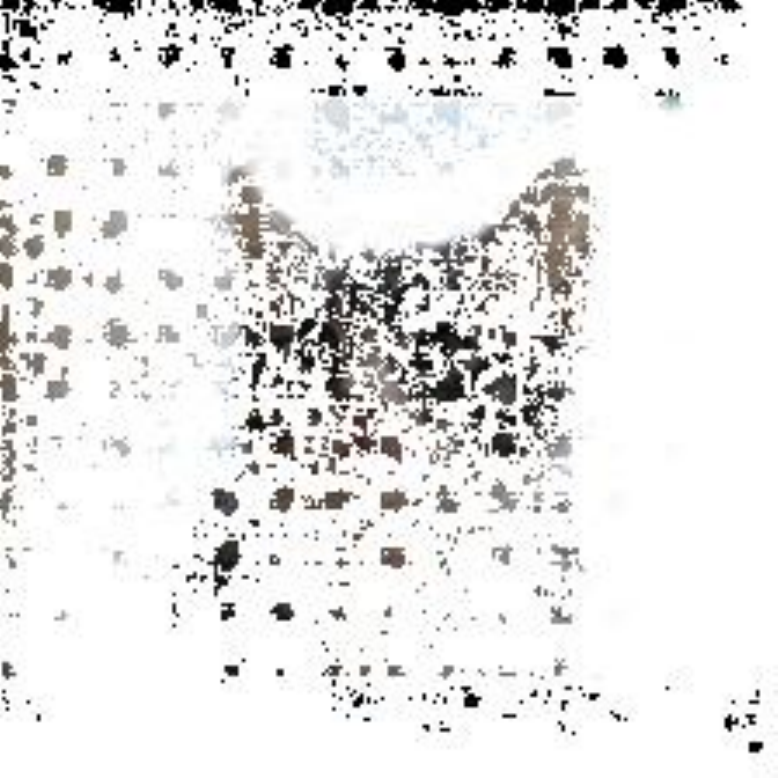} &
 \includegraphics[width=0.62in]{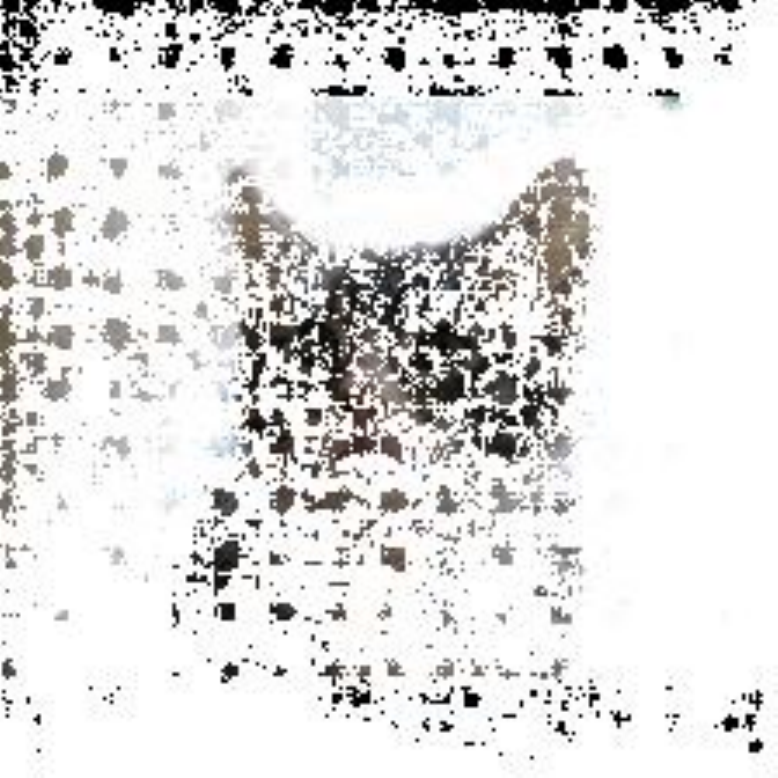} &
 \includegraphics[width=0.62in]{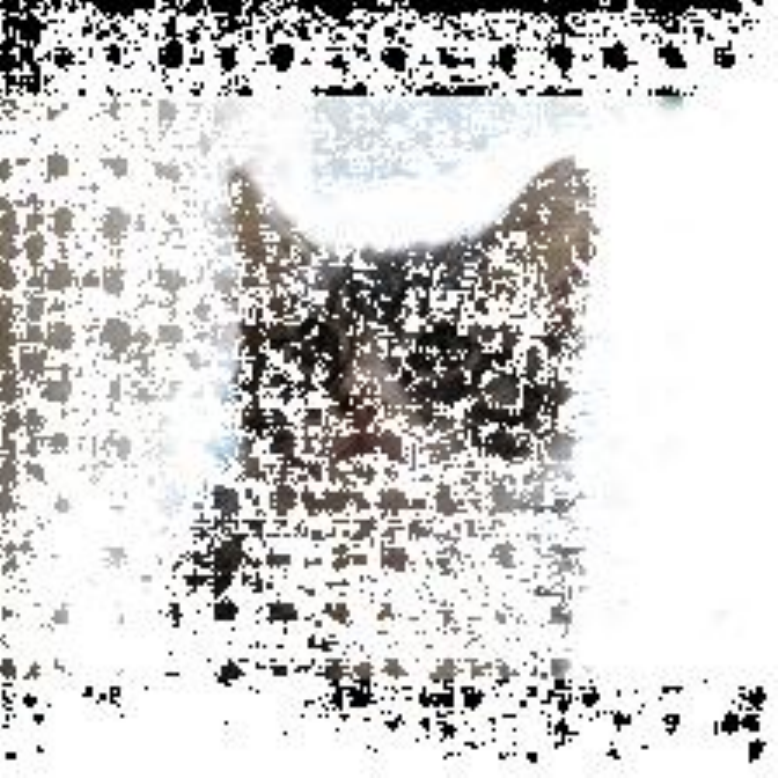} &
 \includegraphics[width=0.62in]{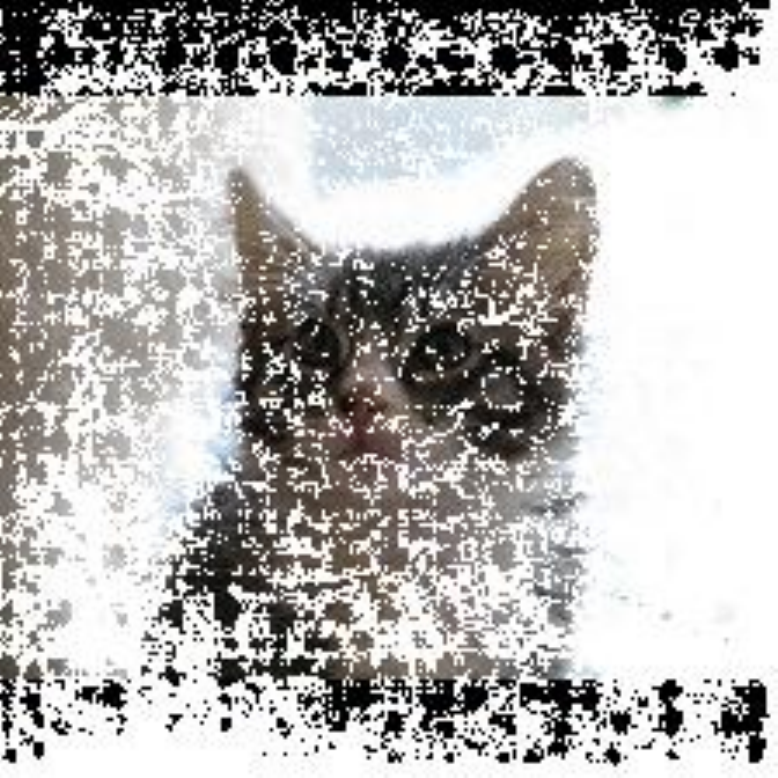} &
 \includegraphics[width=0.62in]{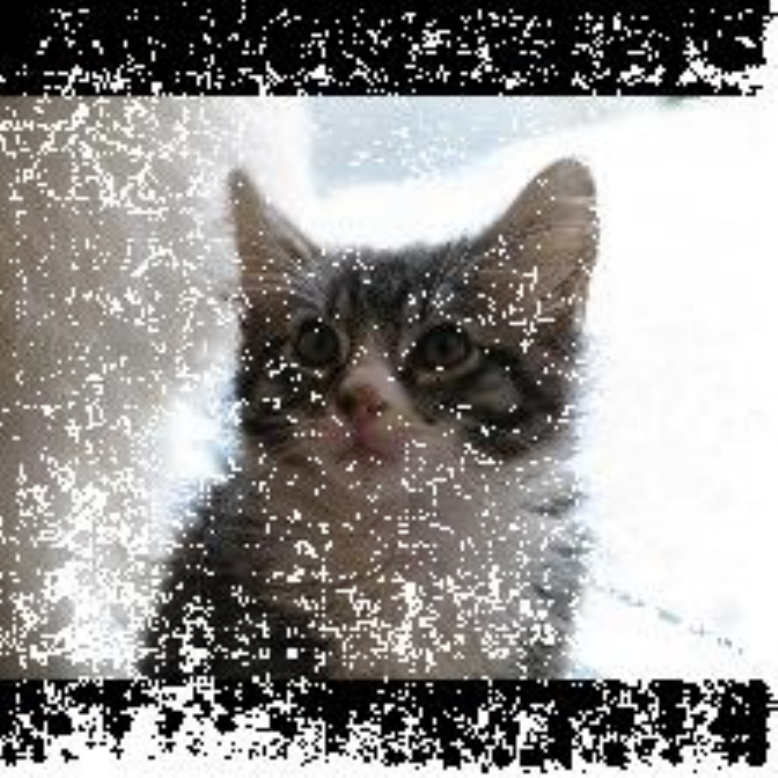}\\
 
  \rotatebox{90}{\textbf{VanillaGrad}} &
 \includegraphics[width=0.62in]{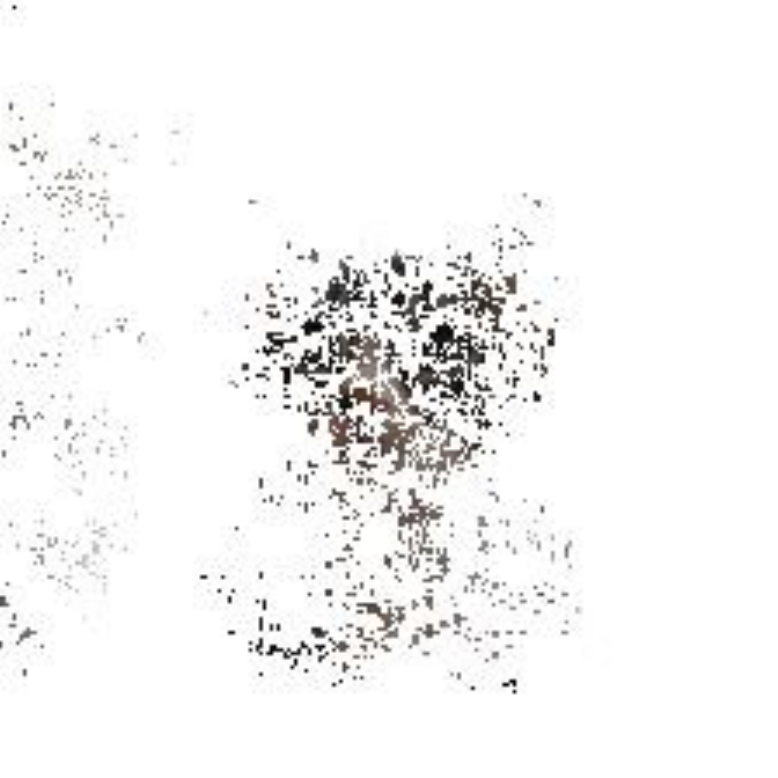} &
 \includegraphics[width=0.62in]{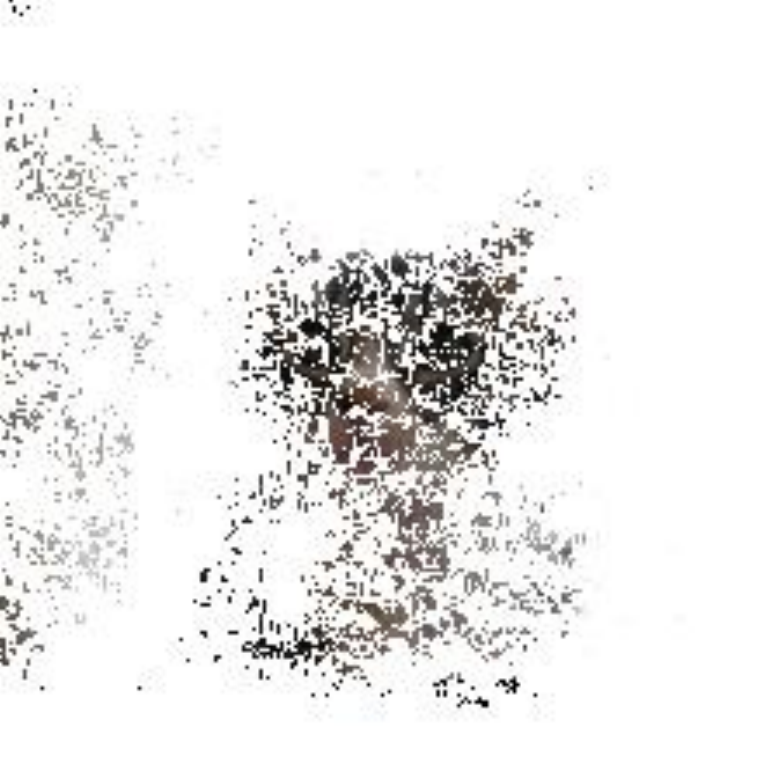} &
 \includegraphics[width=0.62in]{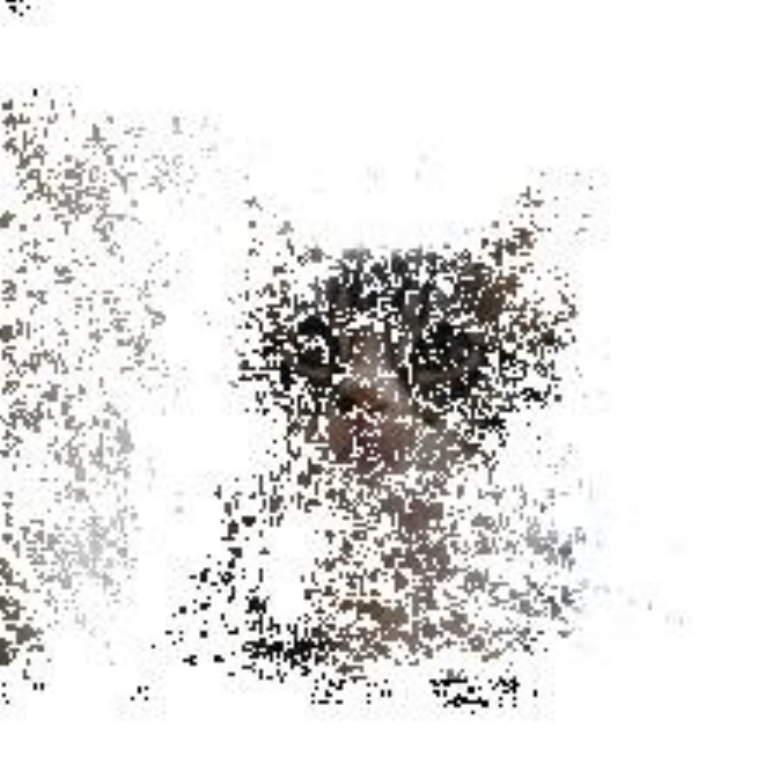} &
 \includegraphics[width=0.62in]{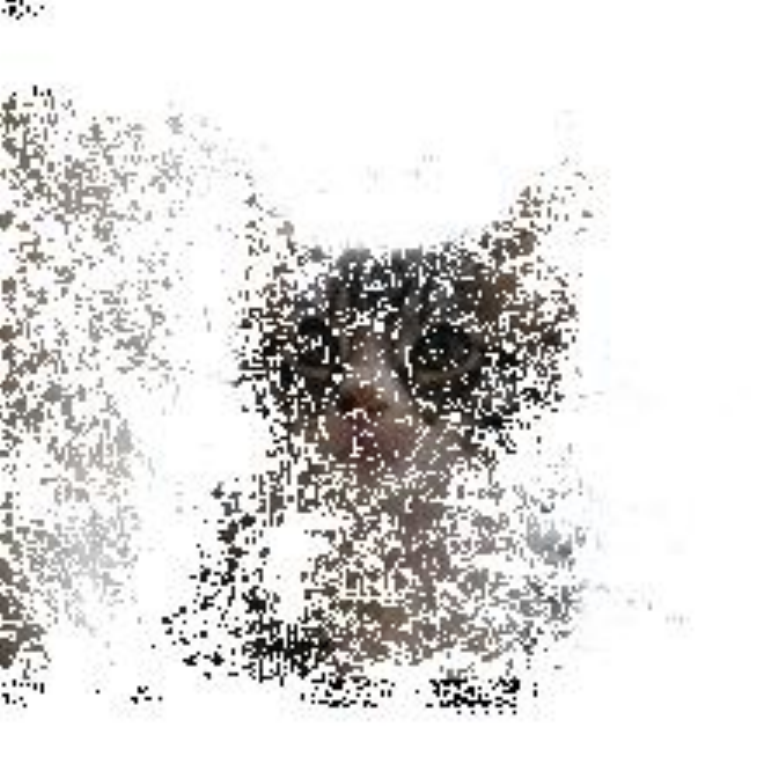} &
 \includegraphics[width=0.62in]{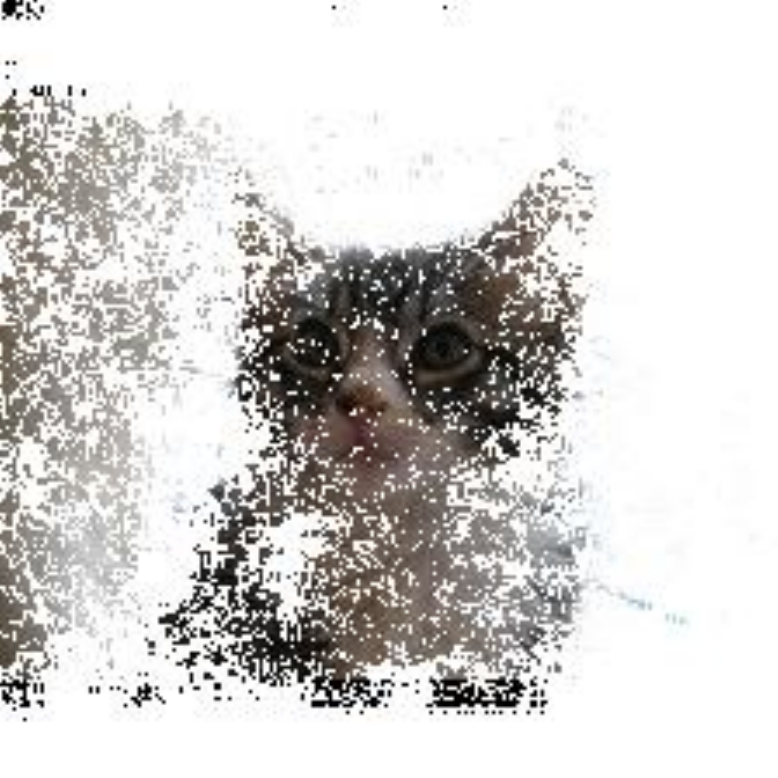} &
 \includegraphics[width=0.62in]{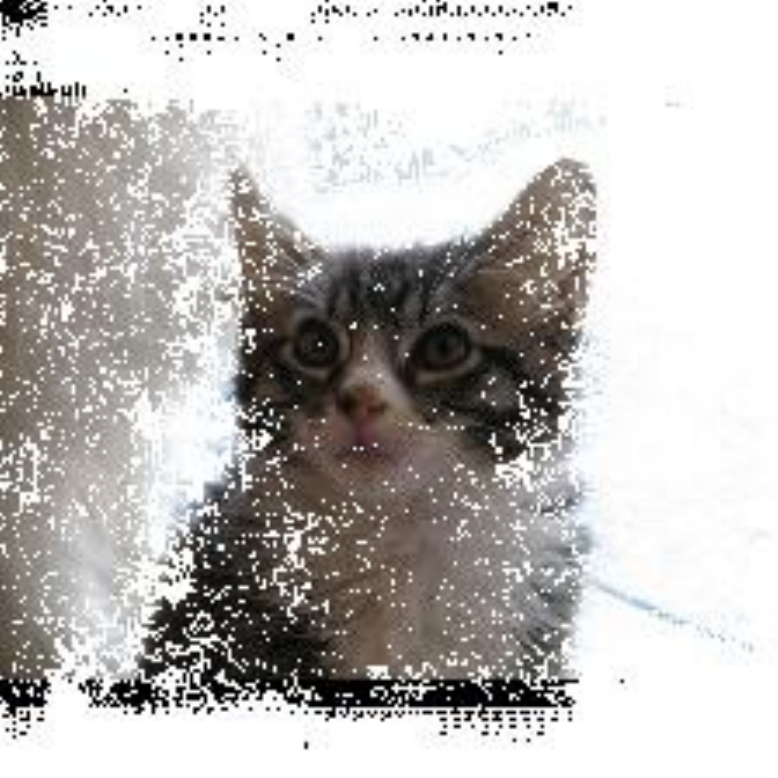} &
 \includegraphics[width=0.62in]{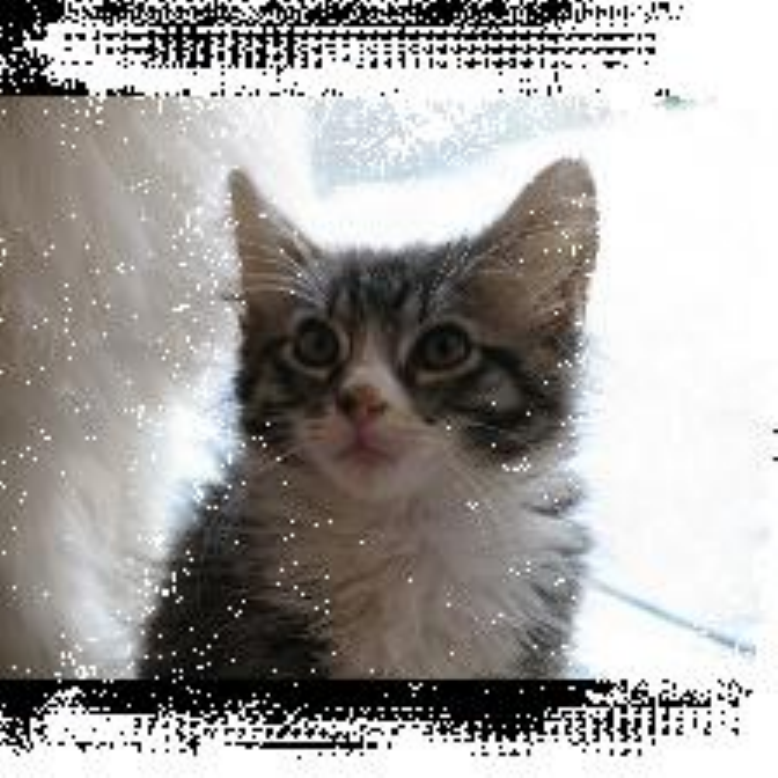}\\
 
  \rotatebox{90}{\textbf{~~Random}} &
 \includegraphics[width=0.62in]{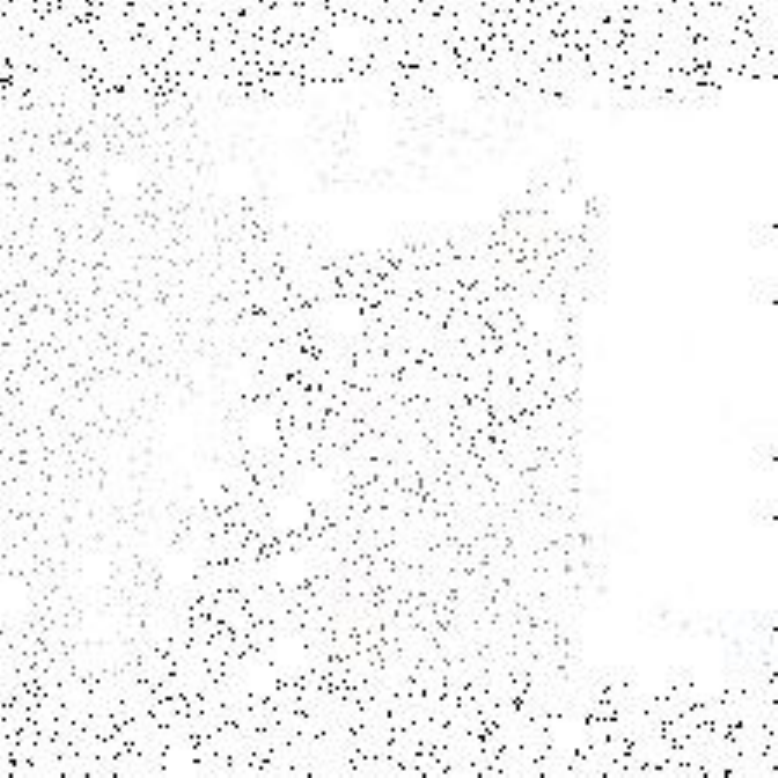} &
 \includegraphics[width=0.62in]{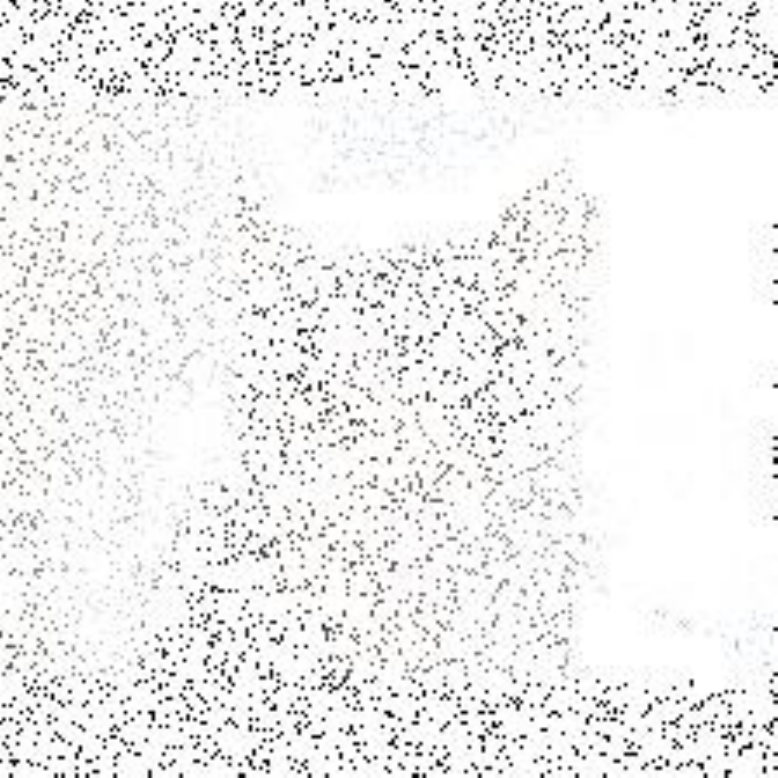} &
 \includegraphics[width=0.62in]{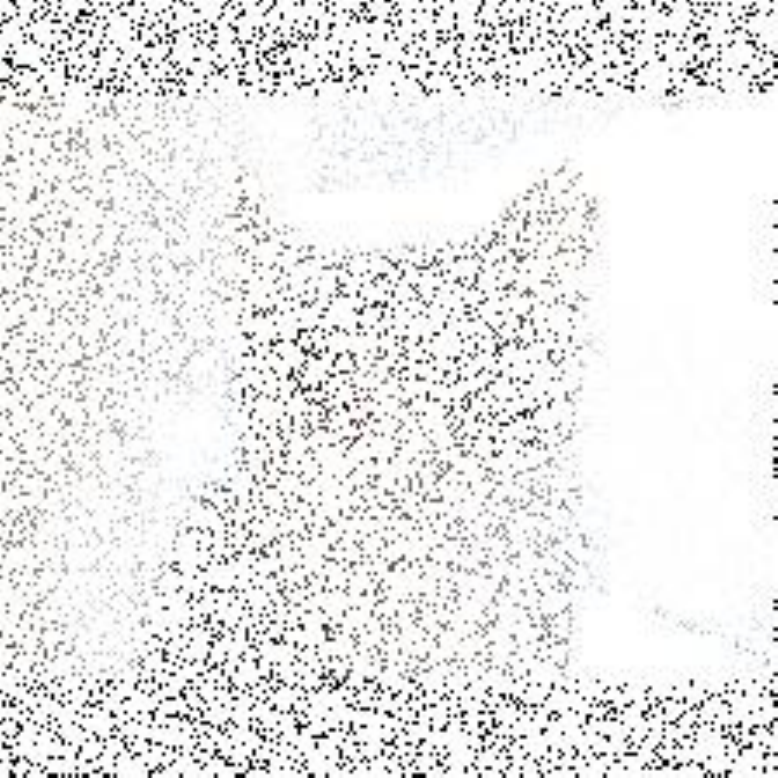} &
 \includegraphics[width=0.62in]{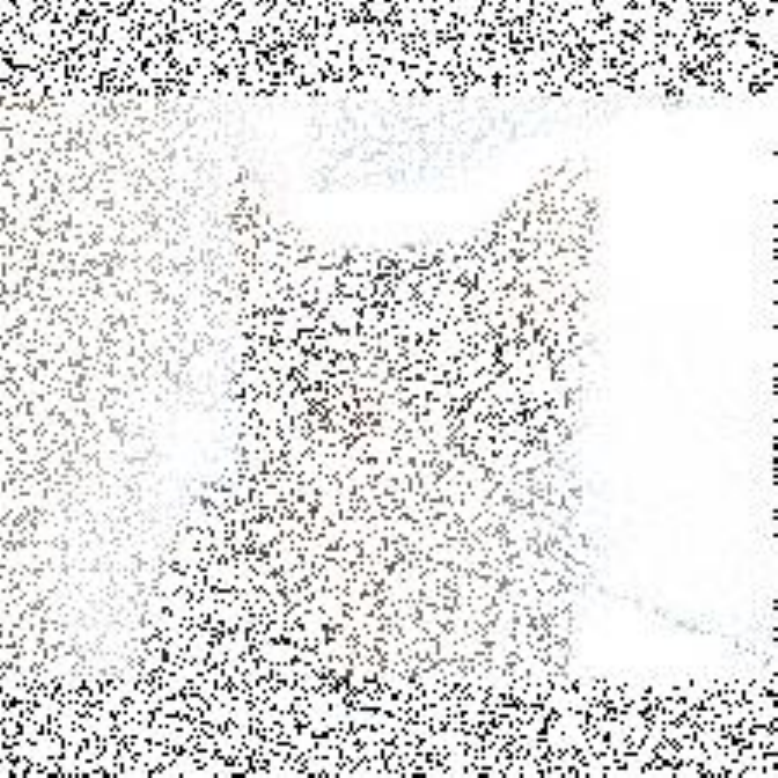} &
 \includegraphics[width=0.62in]{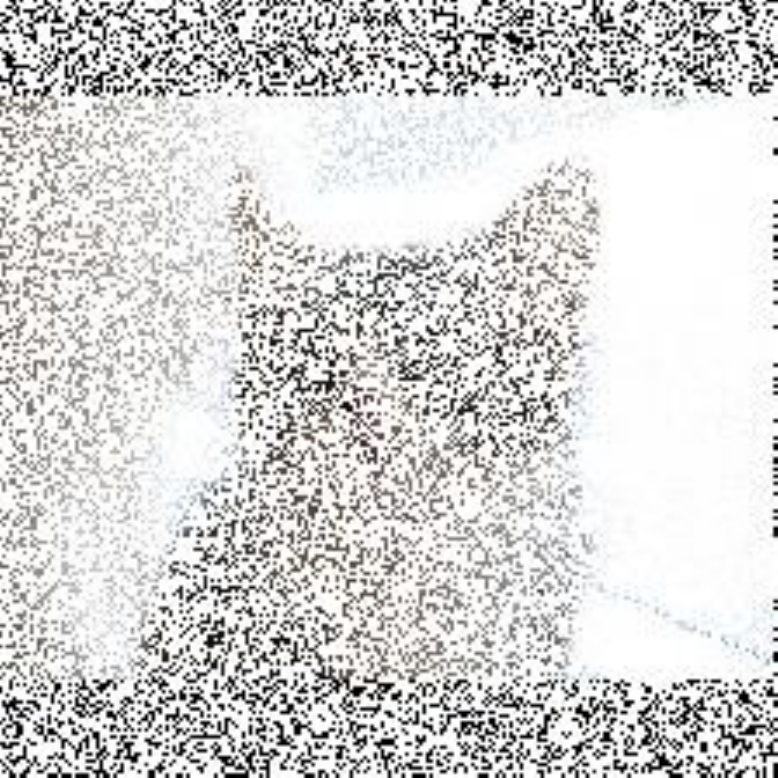} &
 \includegraphics[width=0.62in]{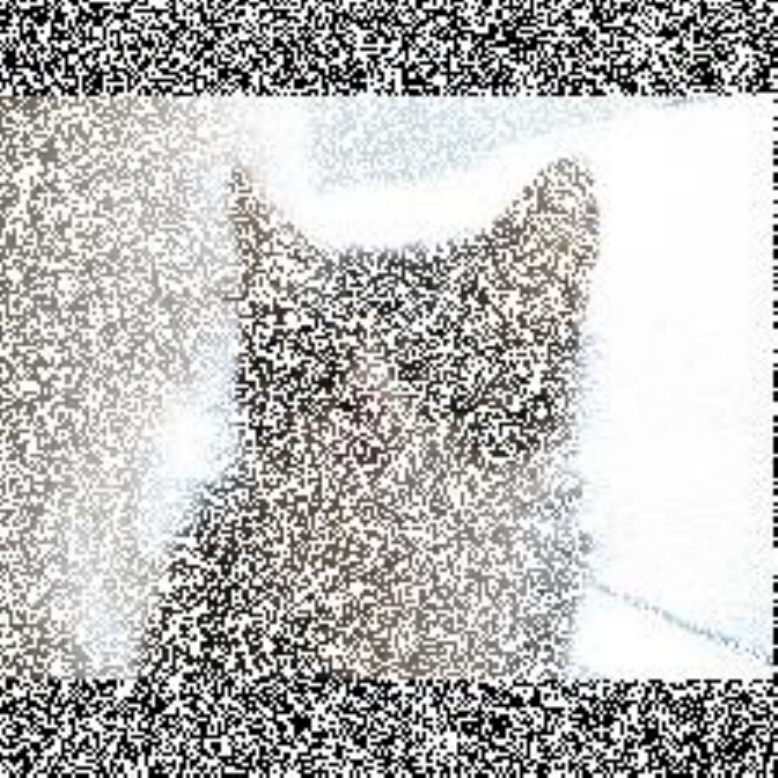} &
 \includegraphics[width=0.62in]{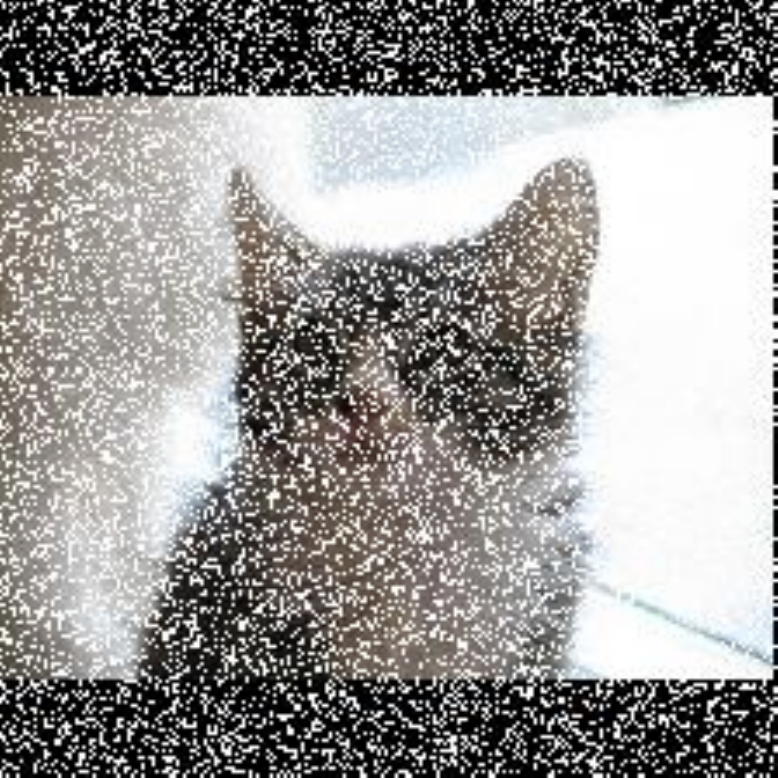}\\
\end{tabular}
    \caption{
    ``Explanations" for a cat image provided by five XAI methods (Grad-CAM, Guided-BP, SmoothGrad, VanillaGrad, and Random) at different exposure rates (5\%, 10\%, 15\%, 20\%, 30\%, 50\%, and 75\%).
    }
     \label{fig:diverseer}
\end{figure}

Specifically, the evaluation procedure consists of the following steps:
\begin{screen}
1) Prepare a pre-trained prediction model, a dataset, and several XAI methods to be evaluated.\\
2) Apply all XAI methods and a random baseline to each image.\\
3) Get a saliency map from each XAI method and the random baseline. \\
4) A series of images with a part of top importance pixel features are generated from the saliency map (Figure \ref{fig:diverseer}).\\
5) Start with the smallest percentage of pixels (e.g., 5\%) and ask the crowd workers about the class of object. If they do not know, show more pixels and record the percentage of images when the worker answers correctly.
\end{screen}

%% file: result.tex
\section{Results}
We conduct experiments to answer the following four questions:\\ 
Q1. Are human and automated evaluations really different? \\
Q2. Which XAI methods are deemed better by humans? \\
Q3. Which automated evaluation scheme is closer to humans? \\
Q4. How does the number and ability of crowd evaluators affect the evaluation results?\\

\subsection{Experimental settings}
We use two datasets for the evaluation; namely, Food101~\cite{bossard14} and Animal95. 
Animal95 is a subset of OpenImages v6 dataset~\cite{OpenImages}.
We use bounding box data to filter the dataset to extract object areas.
Subsequently, we select only images with either single animals or multiple animals of the same class.
Images with multiple classes (multiple types of animals) are not included, but those with several animals of a single class (e.g., three dogs) or non-animal classes are included.

We shortlist the 30 most common food classes from all the classes in Food101, and randomly select ten images in each class.
Similarly, we select 95 most frequent animals from OpenImages for the Animal95 dataset.

%\subsection{Task description}

The XAI methods used in our experiments include GradCAM, Guided-backpropagation, SmoothGrad, and Vanilla Gradient. 
We implement the four XAI methods as well as the random baseline with a pre-trained ResNet50 model~\cite{he2016deep}. 
In total, 1500 pairs and 4750 of (image, XAI method) are generated for Food101 and Animal95, respectively\footnote[2]{$30 \text{ classes} \times 10 \text{ images} \times 5 \text{ XAI methods} = 1500$ for Food101, and $95 \text{ classes} \times 10 \text{ images} \times 5 \text{ XAI methods} = 4750$ for Animal95.}.

We use Amazon Mechanical Turk (AMT)\footnote[3]{\url{https://www.mturk.com/}} and Lancers\footnote[4]{\url{https://www.lancers.jp/}} as the crowdsourcing platforms for crowd-based evaluation.
Each (image, XAI method)-pair is evaluated ten times. 
Each crowd worker is required to evaluate 20 pairs for a reward of USD 0.5 in AMT, or JPY 40 in Lancers. To avoid biases, we randomly sample (image, XAI method)-pairs assigned to each worker. 
Approximately 3200 crowd workers participate in the evaluation tasks.

We compare ROAR, KAR, ROAE, and KAE schemes (introduced in~\ref{subsec:xai_methods}) as the representatives of automated evaluation schemes.
The performance of each XAI method is evaluated by an accuracy-exposure curve.
For the crowd-based evaluation, We calculate the average accuracy of crowd answers at each exposure rate, while the human evaluators are replaced by a machine classifier in the automated evaluation.

In addition, we also provide the area under curve (AUC) of each accuracy-exposure curve. 
Let a series of exposure rates be $r_1=0,r_2,r_3, \ldots, r_n=1$, where $r_i<r_j$ for  $i<j$.
Let $a_{i}^k$ denote the accuracy at exposure rate $r_i$ for XAI method $k$ both in crowd-based and automated evaluations. 
The value of AUC in XAI method $k$ denoted by $\text{AUC}^k $ is defined as
$$ \text{AUC}^k=\sum_{i=2}^{n} \frac{1}{2} (r_i-r_{i-1})(a_{i}^k+a_{i-1}^k). $$

\subsection{Results}

\subsubsection{Q1. Are human and automated evaluations really different? } \hfill\\
The first question we investigate is the difference between automated and crowd-based evaluation schemes, because the latter requires higher time and financial costs, and there is no reason to resort to human evaluation if they both give the same results.

\begin{figure}[!tb]
    \centering
         \includegraphics[width=4.8in]{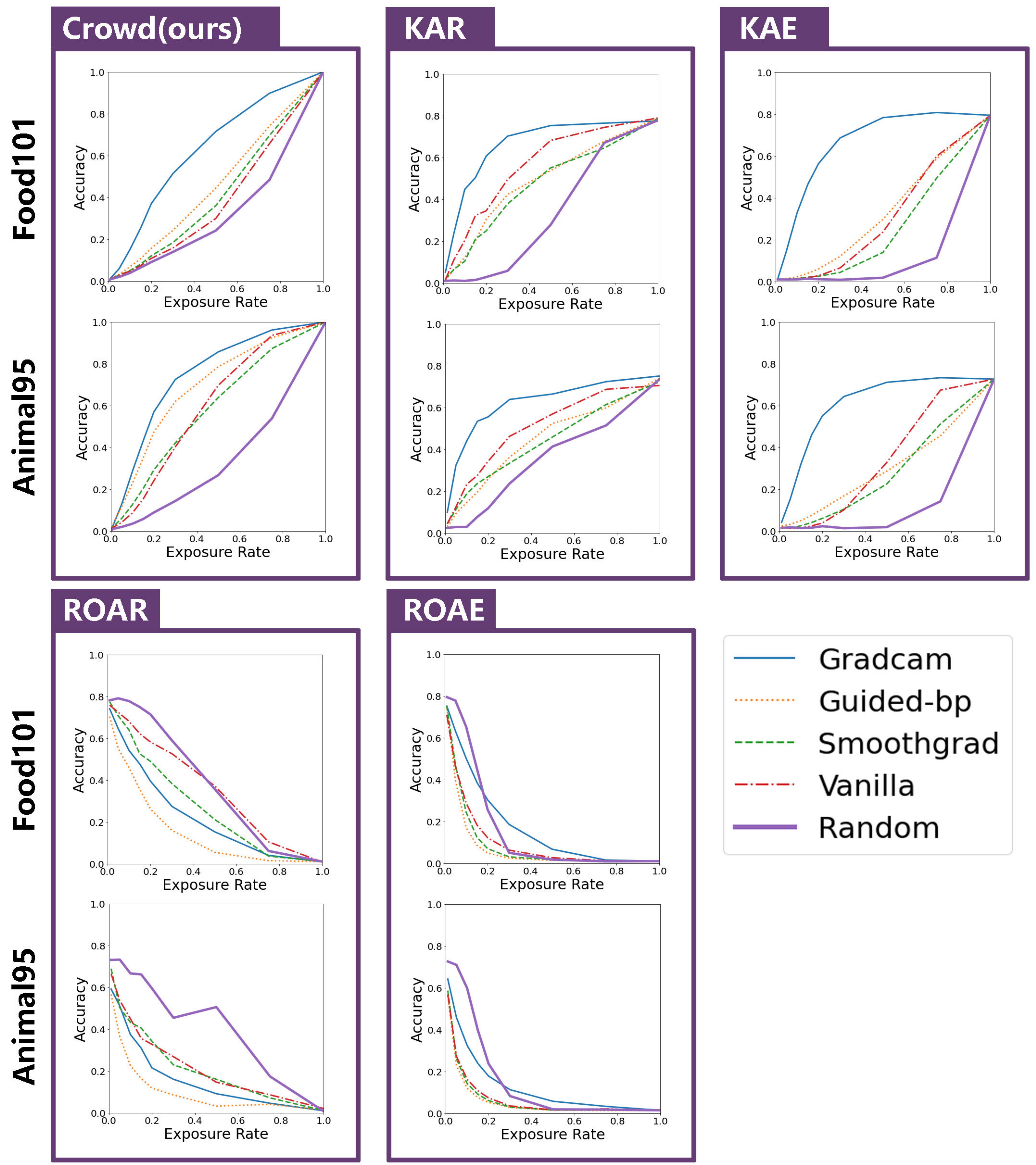}
    \caption{
 Performance curves of different XAI methods on the two datasets (Food101 and Animal95). The horizontal and vertical axis indicates the exposure rate and accuracy, respectively. For the crowd-based evaluation (Crowd), KAR, and KAE, upper curves indicate better performance. In contrast, for ROAR and ROAE, lower curves are better.
    }
    \label{fig:resultcurve}
\end{figure}

Figure~\ref{fig:resultcurve} shows the accuracy-exposure curves of different XAI methods at different exposure rates by different evaluation methods.
For the crowd-based evaluation scheme (denoted by Crowd) and two automated evaluation schemes (KAR and KAE), higher curves indicate better performance. 
In contrast, for the other two automated evaluation schemes, ROAR and ROAE, lower curves indicate better performance.

Table~\ref{tab:AUC_Results} shows the AUCs of each scheme; each row and column correspond to an evaluation scheme and an XAI method, respectively.
The numbers in the brackets show the ranks. 
The bold numbers show the best results. Although we can see some consistency between the ranking of AUC by the crowd-based ranking and those by the automatic evaluation schemes, no automated evaluation scheme obtains the same ranking of AUC as the crowd-based evaluation. (We will see more detailed comparisons later.)

\begin{figure}[!tb]
    \centering
         \subfloat[Image difficulty (Food101)]{
            \includegraphics[width=2.2in]{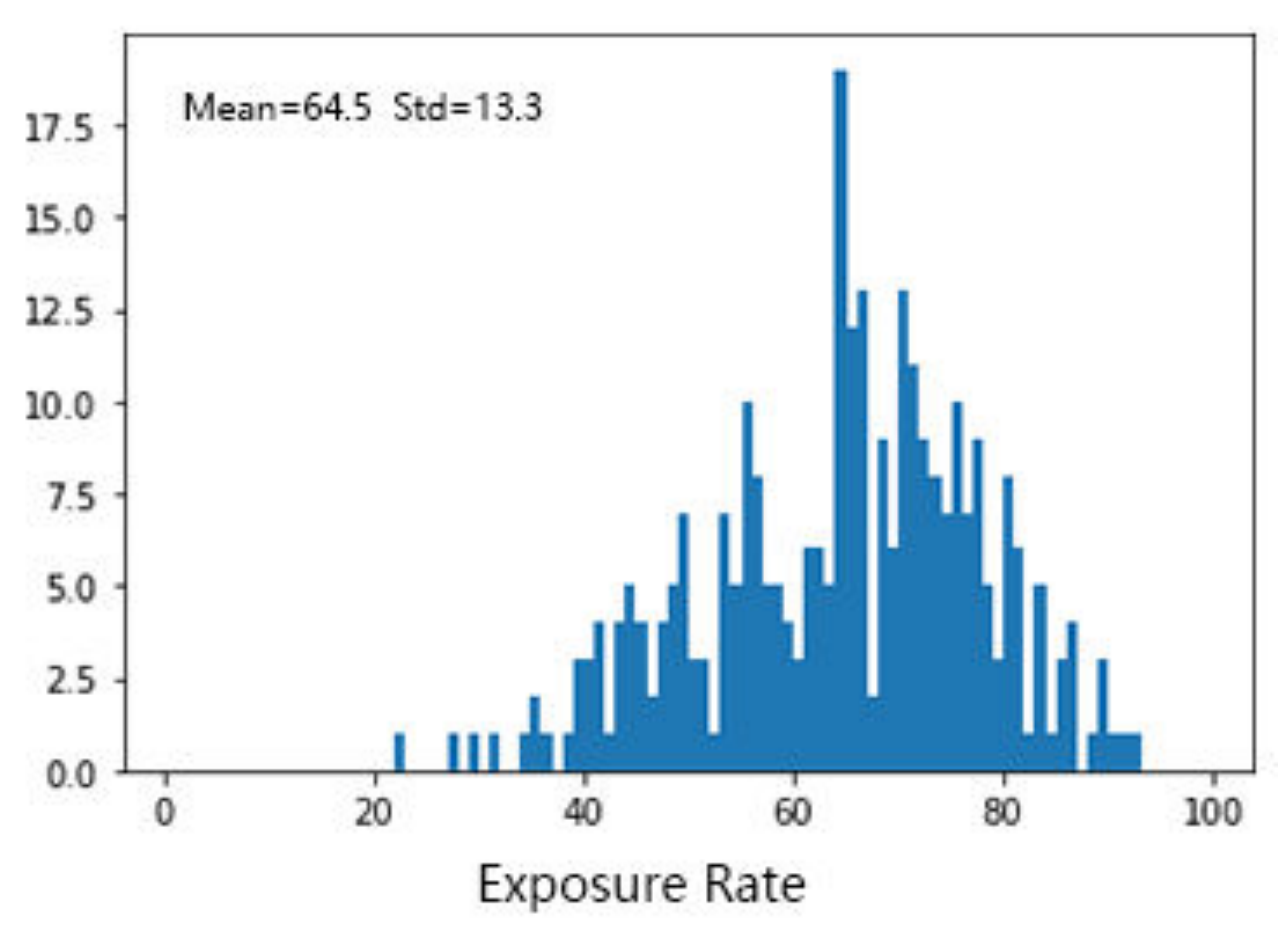}%
            \label{Food101imageAccuracyfre}} 
            \hspace{5mm}
         \subfloat[Image difficulty (Animal95)]{
            \includegraphics[width=2.2in]{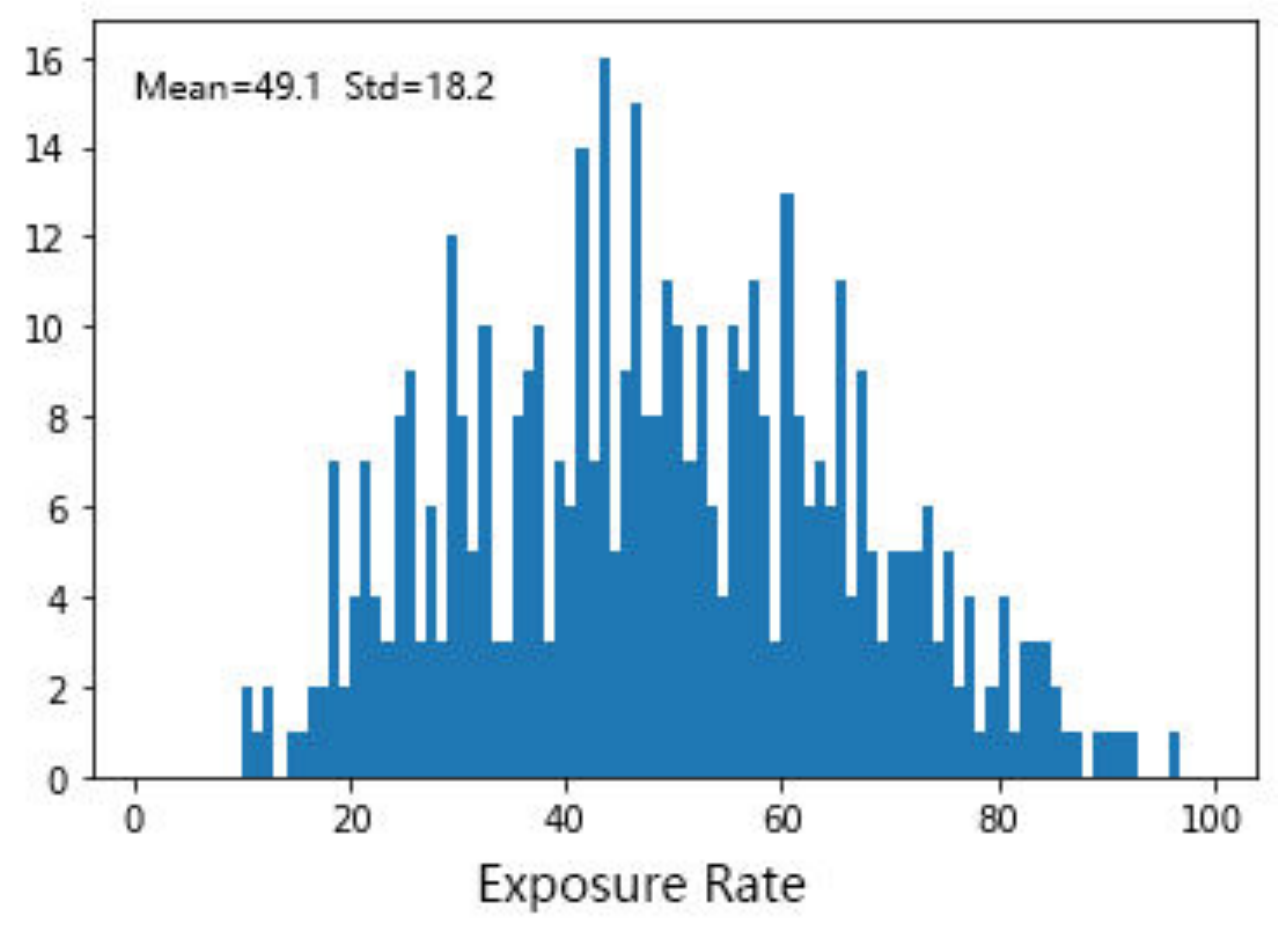}%
            \label{Animal95imageAccuracyfre}} \\
        
        \subfloat[Worker ability (Food101)]{
            \includegraphics[width=2.2in]{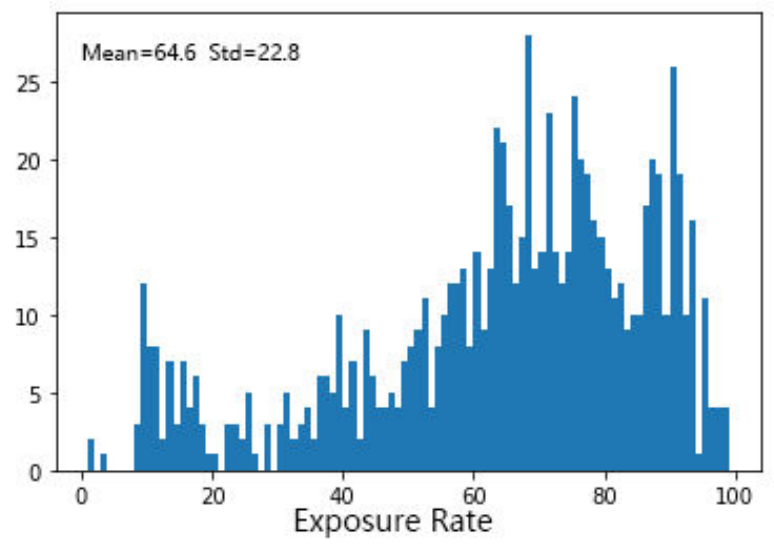}%
            \label{Food101Accuracyfre}} 
            \hspace{5mm}
         \subfloat[Worker ability (Animal95)]{
            \includegraphics[width=2.2in]{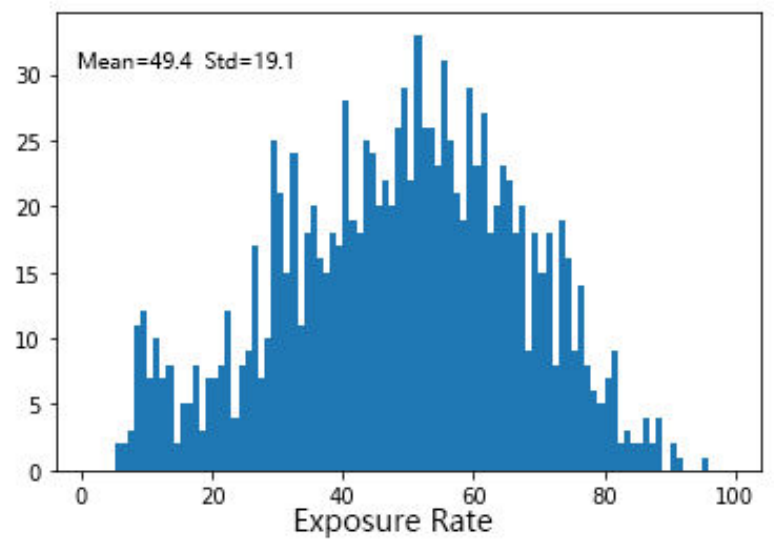}%
            \label{Animal95Accuracyfre}}   
            
    \caption{
      Histograms of average exposure rates of correct answers. The horizontal and vertical axes indicate the exposure rate and frequency, respectively; the top and bottom rows indicate the frequencies of images and workers, which show the distributions of ``image difficulty" and ``worker ability", respectively. 
      Comparing (a) and (b), the mean image difficulty of Animal95 dataset is higher than that of Food101, indicating the Animal95 dataset is relatively easier than Food101. 
      In the bottom row ((c) and (d)), the variance of the worker ability in the Food 101 dataset is higher than that of Animal95, which is probably because the difficulty of recognizing food can be significantly affected by cultural differences. 
      In spite of the large variations in the worker ability, Table~\ref{tab:worknum} shows they have  no significant impacts on the results.
    }
    \label{fig:imageaccuracyfre}
\end{figure}

\begin{figure}[!tb]
    \centering
         \subfloat[Spearman@Food101]{
            \includegraphics[width=2.2in]{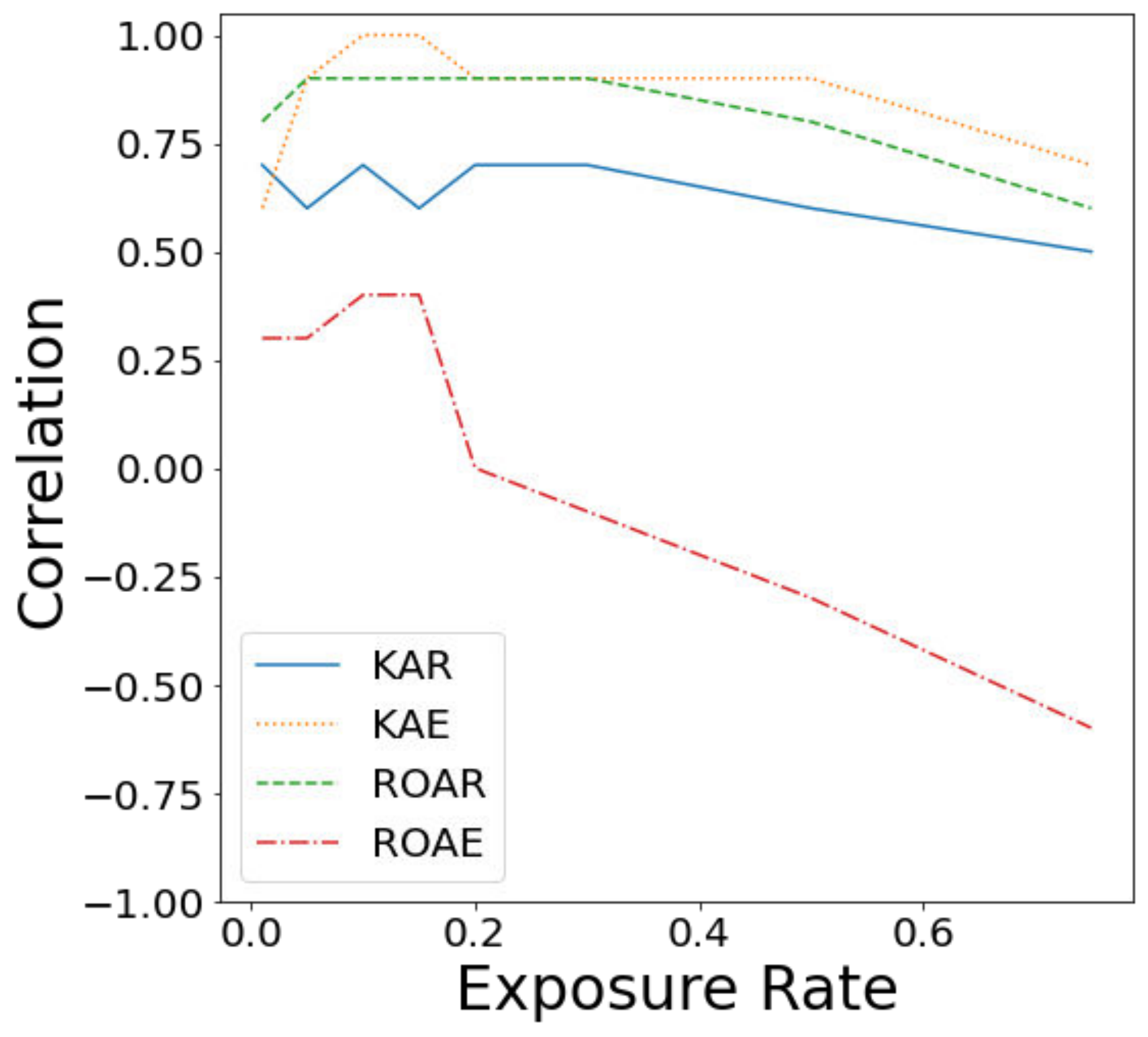}%
            } 
            \vspace{2mm}
         \subfloat[Kendall@Food101]{
            \includegraphics[width=2.2in]{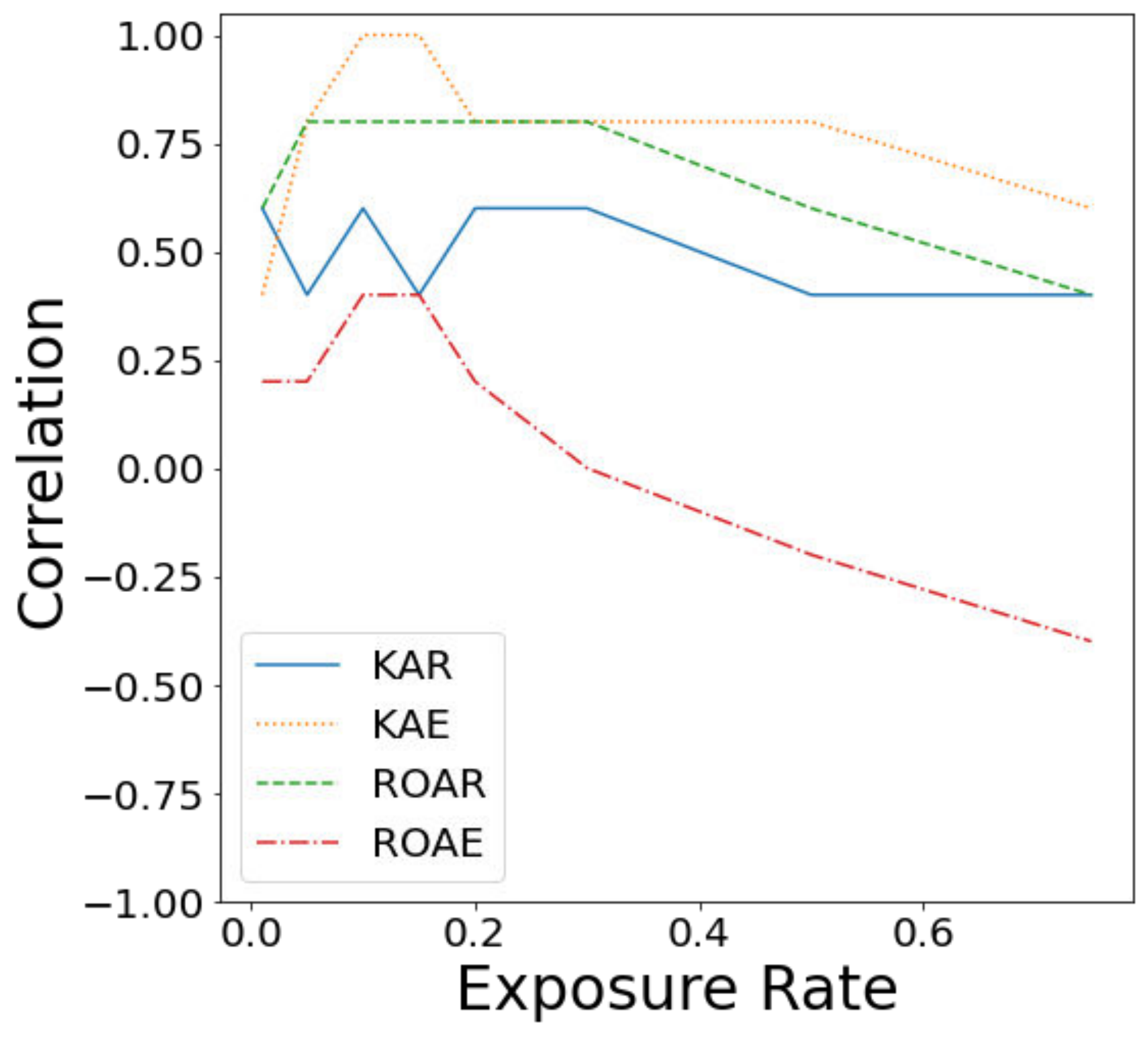}%
           } \\   
        \subfloat[Spearman@Animal95]{
            \includegraphics[width=2.2in]{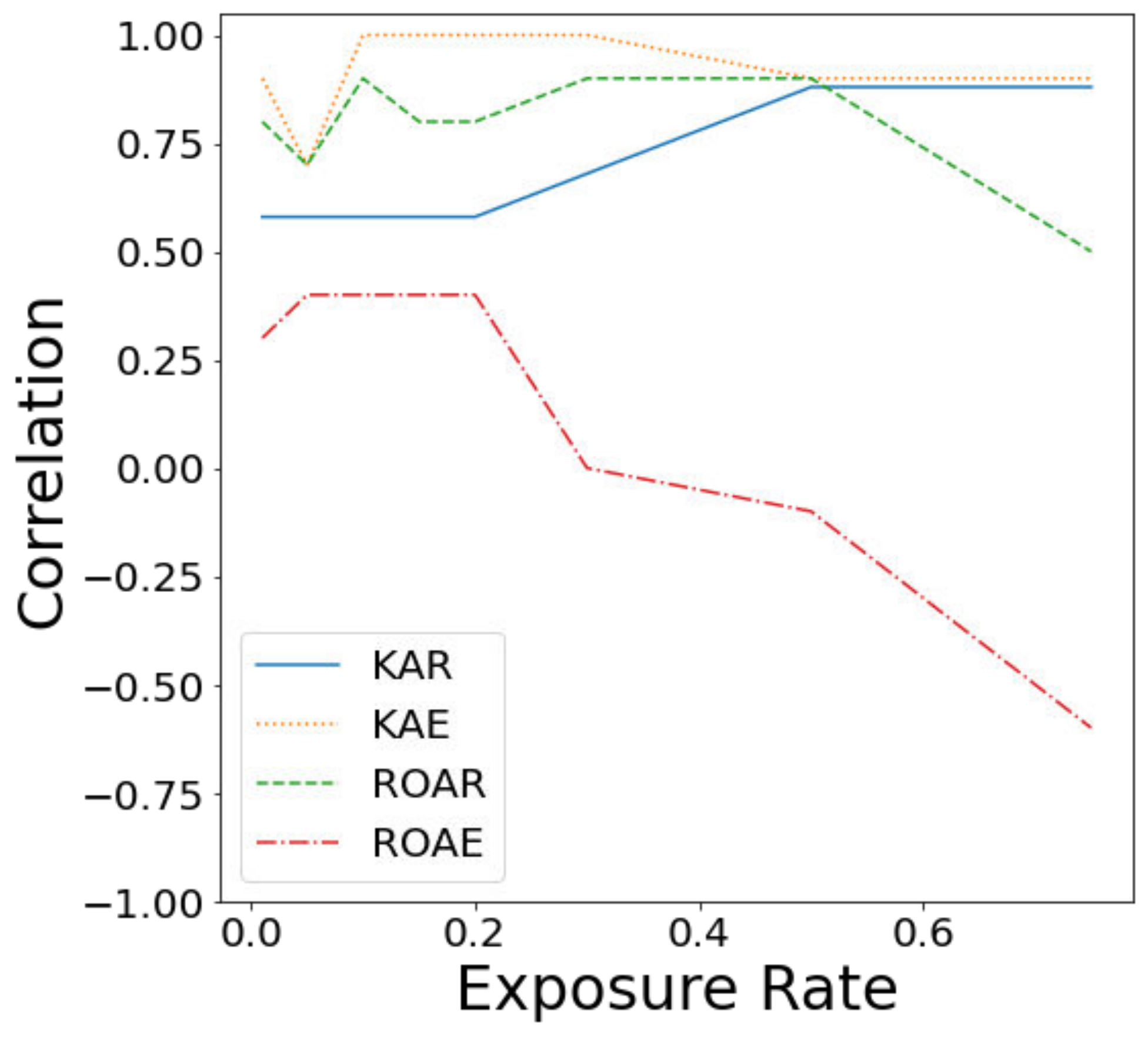}%
            } 
            \vspace{2mm}
         \subfloat[Kendall@Animal95]{
            \includegraphics[width=2.2in]{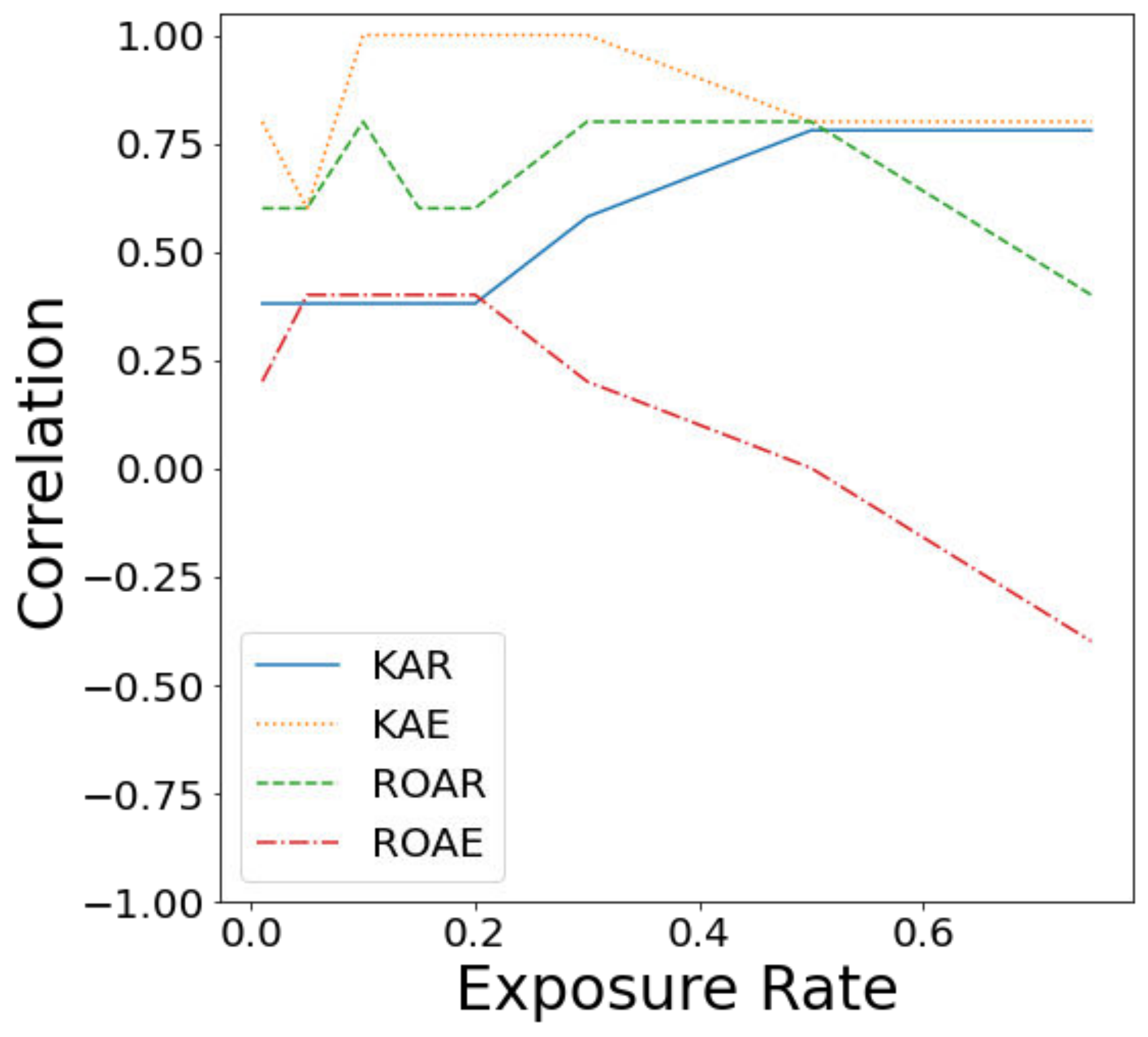}%
            }  
    \caption{
      Correlations of XAI methods ranking between the crowd-based evaluation and the four automated evaluations at different exposure rates in the Food101 and Animal95 datasets (the higher, the better). The correlations are given both in the Spearman ranking correlation and the Kendall ranking correlation.
      The horizontal and vertical axes indicate the exposure rate and the ranking correlation value, respectively.
    }
    \label{fig:rankingWithFixedRate}
\end{figure}

\begin{table}[!tb]
\centering
\caption{AUCs of different methods by different evaluation schemes. The numbers in the bracket show the rank of the XAI method. The bold numbers show the best results. For the crowd-based evaluation (Crowd), KAR, and KAE, larger AUC values show better performance, that is, important areas of images are shown earlier. On the other hand, for ROAR and ROAE, smaller AUCs indicate better performance, that is, important areas are removed earlier. (Also see Figure~\ref{fig:resultcurve}.) 
}
\begin{tabular}{|c|c|c|c|c|c|c|} \hline
Dataset & Scheme &GradCAM&Guided-bp&SmoothGrad&Vanilla Gradients & Random \\ 
\hline
\multirow{5}*{Food101} & Crowd (ours)& \textbf{0.639} (1) &0.469 (2)&0.425 (3)&0.396 (4)&0.334 (5) \\  \cline{2-7}

~ & KAR& \textbf{0.667} (1) & 0.494 (3)& 0.478 (4)& 0.570 (2) & 0.340 (5) \\   \cline{2-7}

 ~ & KAE& \textbf{0.669} (1) & 0.340 (2) & 0.265 (4) & 0.316 (3) & 0.136 (5) \\  \cline{2-7}

~& ROAR & 0.211 (2) & \textbf{0.140} (1) &0.258 (3) &0.346 (4)&0.366 (5) \\   \cline{2-7}

~& ROAE& 0.159 (5) & \textbf{0.060} (1) & 0.072 (2) & 0.087 (3) & 0.140 (4) \\   \cline{1-7}

\multirow{5}*{Animal95} & Crowd (ours)& \textbf{0.752} (1)&0.696 (2)&0.592 (4)&0.608 (3)&0.354 (5) \\  \cline{2-7}

~ & KAR  & \textbf{0.627} (1) & 0.456 (3) & 0.445 (4) & 0.515 (2) & 0.365 (5) \\  \cline{2-7}

~ &KAE& \textbf{0.619} (1) & 0.311 (3) & 0.294 (4) & 0.354 (2) & 0.137 (5) \\  \cline{2-7}

~ &ROAR & 0.142 (2) & \textbf{0.088} (1) & 0.194 (3) & 0.200 (4) & 0.385 (5) \\  \cline{2-7}

~ &ROAE& 0.115 (4) & \textbf{0.048} (1) & 0.054 (2) & 0.059 (3) & 0.137 (5) \\ \cline{1-7}

\end{tabular}
\label{tab:AUC_Results}
\end{table}

Now, we discuss the impact of different datasets on the results.
It is evident from Table~\ref{tab:AUC_Results} that the values of AUC are different among the two datasets, Food101 and Animal95. The AUCs for Animal95 datasets are generally better than those for Food101 in all the schemes. This is probably because it is rather easier to recognize animals than foods; this is also suggested by Figure~\ref{fig:imageaccuracyfre} showing the distribution of the ``difficulty" of the images.

In contrast, Figure~\ref{fig:resultcurve} shows that the ranking of XAI methods is not entirely different among the datasets, except for the slight difference in SmoothGrad and Vanilla Gradients.
This shows that the difference of datasets does not significantly impact the relative superiority or inferiority of the different schemes.
However, this conclusion is drawn from only two datasets and needs to be validated with more datasets in the future.

\subsubsection{Q2. Which XAI methods are deemed better by humans?} \hfill\\
Because different XAI methods provide different pixel rankings, the next question we investigate is which XAI method is more reliable. It can be observed from Table~\ref{tab:AUC_Results} that GradCAM exhibits the best performance in the crowd-based evaluation scheme.
This is probably because GradCAM produces low resolution feature maps (e.g., $7\times 7$ for ResNet family when the input size is $224 \times 224$), which are then linearly interpolated to the resolution of the input image, thereby producing mostly connected regions rather than distributed regions. 

In addition, this result is also consistent with the conclusion of previous work on sanity checking~\cite{adebayo2018sanity}.

\subsubsection{Q3. Which automated evaluation scheme is closer to humans?}\hfill\\
Once we assume that the human assessments are the ground truths, automated evaluation scheme which is closer to the crowd-based evaluation scheme indicates better performance.
Quantifying the goodness of automated evaluation schemes is not only useful for evaluating them but will also help improving themselves.

We investigate the ranking similarity between the four automated evaluations and the crowd-based evaluations.
Figure~\ref{fig:rankingWithFixedRate} shows the correlations of XAI methods ranking between the crowd-based evaluation and the four automated evaluations at different exposure rates in the Food101 and Animal95 datasets.

Most of the automated evaluation schemes show positive correlations with crowd-based evaluation, but only ROAE shows the lowest correlations, and even gives negative correlations for high exposure rates; this is probably because the mechanism of ROAE equals adding white noises to the images at high exposure rates, which is also known as an approach for generating adversarial examples.

ROAR shows the better performance than ROAE, which is consistent with the report in the previous study~\cite{hooker2019benchmark}, which implies the importance of re-training.
KAE consistently performs well independent of the change of datasets, correlation types, and exposure rates.
KAR performs sub-optimally, but it maintains almost the same performances at high exposure rates, while other automated evaluation schemes tend to decrease the performance at high exposure rates.

\subsubsection{Q4. How does the number and ability of crowd evaluators affect the evaluation results?  }\hfill\\
Finally, we investigate the stability of the proposed crowd-based evaluation scheme in terms of the number of crowd workers participating in the evaluation.
Crowd workers have different abilities and diligence; for example, some crowd workers do not work seriously on tasks.  
Figure~\ref{Food101Accuracyfre} and Figure~\ref{Animal95Accuracyfre} show the histograms of the average exposure rate at which each crowd worker made a correct answer, which can be considered as the distribution of the worker ability; large variations are observed in the ability of the workers.

Table~\ref{tab:worknum} summarizes the AUC values of the performance curves when the average number of workers per (XAI method, image)-pair is changed.
Some variations are observed in the results when the average number of workers was changed; however, no significant change was found in the qualitative results, which shows the stability and efficiency of the proposed crowd evaluation scheme.

\begin{table}[!tb]
\centering
\caption{AUCs for different average numbers of crowd workers per (image, XAI method)-pair.
The numbers in the bracket show the rank of the XAI method. Although the performance of crowd workers varies greatly due to their different ability and diligence (Figure~\ref{fig:imageaccuracyfre} (c)(d)), the ranking of XAI methods does not change according to the number of workers. %(See also Figure~\ref{fig:worknum}). 
}
\begin{tabular}{|c|c|c|c|c|c|c|} \hline

Dataset&Workers per image&GradCAM&Guided-bp&SmoothGrad&Vanilla Gradients & Random \\
\hline
\multirow{7}*{Food101} & 0.3& 0.647 (1) & 0.492 (2) & 0.474 (3) & 0.422 (4) & 0.337 (5)\\ \cline{2-7}

~& 0.5& 0.618 (1) & 0.468 (2) & 0.450 (3) & 0.419 (4) & 0.348 (5) \\ \cline{2-7}
~&1& 0.610 (1) & 0.465 (2) & 0.446 (3) & 0.422 (4) & 0.331 (5) \\  \cline{2-7}
~&3& 0.632 (1) & 0.462 (2) & 0.421 (3) & 0.395 (4) & 0.330 (5) \\  \cline{2-7}
~&5& 0.636 (1) & 0.470 (2) & 0.431 (3) & 0.399 (4) & 0.332 (5) \\  \cline{2-7}
~&7& 0.638 (1) & 0.469 (2) & 0.428 (3) & 0.397 (4) & 0.333 (5) \\  \cline{2-7}
~&10& 0.639 (1) & 0.469 (2) & 0.425 (3) & 0.396 (4) & 0.334 (5) \\  \hline
\multirow{7}*{Animal95}&0.3& 0.742 (1) & 0.675 (2) & 0.624 (3) & 0.622 (4) & 0.367 (5) \\  \cline{2-7}
~&0.5& 0.752 (1) & 0.667 (2) & 0.591 (4) & 0.616 (3) & 0.347 (5) \\  \cline{2-7}
~&1& 0.745 (1) & 0.683 (2) & 0.603 (4) & 0.622 (3) & 0.339 (5) \\  \cline{2-7}
~&3& 0.758 (1) & 0.695 (2) & 0.604 (4) & 0.608 (3) & 0.365 (5) \\  \cline{2-7}
~&5& 0.755 (1) & 0.696 (2) & 0.596 (4) & 0.608 (3) & 0.357 (5) \\  \cline{2-7}
~&7& 0.752 (1) & 0.696 (2) & 0.593 (4) & 0.609 (3) & 0.358 (5) \\  \cline{2-7}
~&10& 0.752 (1) & 0.696 (2) & 0.592 (4) & 0.608 (3) & 0.354 (5) \\  \hline
\end{tabular}
\label{tab:worknum}
\end{table}

%% file: conclusion.tex
\section{Conclusion}
In this study, we investigated schemes for evaluation of XAI methods.
Based on the hypothesis that interpretability for humans can ultimately only be assessed by humans, We proposed a new human-based evaluation scheme using crowdsourcing and compared it with existing automated evaluation schemes.
We convened a total of 3,200 crowd workers to conduct experiments using four XAI methods and two datasets.
The results showed that there are differences between the crowd-based evaluation and automatic evaluation.
Among the various automatic evaluation schemes, KAE gave the most similar XAI evaluations to human evaluation.

In the report by Hooker et al.~\cite{hooker2019benchmark}, ROAR performed better than KAR, but the results of our experiment indicate the opposite, which can be further investigated in the future. 

In addition, among the four XAI methods, Grad-CAM was found to be the XAI method closest to human evaluation.
This is rather counter intuitive if we focus only on saliency maps because Guided-Backpropagation and SmoothGrad highlight the outline of objects more accurately (as shown in Figure~\ref{fig:catdiffmap}); however, in our scheme, we present a combination of the original image and the saliency map so that Grad-CAM can convey more information with the fewer pixels.

We also confirmed that the number of crowd workers and datasets did not significantly impact the results; however, larger-scale experiments using more datasets will be desirable in the future.